# Numerical and Theoretical Modeling of Droplet Impact on Spherical Surfaces

Hussein N. Dalgamoni[†] and Xin Yong[*]

Department of Mechanical Engineering, Binghamton University, The State University of New York, Binghamton, New York 13902, U.S.A.

[†] Department of Mechanical Engineering, The Hashemite University, Zarqa 13133, Jordan

[*] Email: xyong@binghamton.edu

**Abstract**

Droplet impact on solid surfaces is a fluid phenomenon widely involved in additive manufacturing, heat management, and coating, in which the ability to exert control over the impact dynamics and duration is critical. While past studies have established a comprehensive understanding of the impact on flat substrates, what we know about the impact dynamics on curved solid surfaces is still limited. This work aims to elucidate the physics of droplet impact on spherical surfaces with different Weber numbers (*We*), radii ($R_s$), and surface wettability ($\theta^{\mathrm{eq}}$) using a combination of axisymmetric lattice Boltzmann method (LBM) and theoretical analysis. The model developed in our previous work [H. Dalgamoni and X. Yong, Phys. Rev. E **98**, 13102 (2018)] was extended and modified for simulating the normal impact of droplet on curved substrates in the low Weber number regime (i.e., $We \leq 15$), in which axisymmetric assumption of droplet deformation holds. The LBM simulations show that $We$, $R_s$, and $\theta^{\mathrm{eq}}$ significantly affect the spreading and recoiling of droplet during impact. The parametric studies uncover five outcomes of impact, which range from complete deposition to total rebound. A simulation-predicted phase diagram was constructed and correlated with the total time that the droplet was in contact with the solid. In addition, a theoretical model based on energy budget during impact was developed to predict the rebound threshold for impact on spherical targets when varying $We$, $R_s$ and $\theta^{\mathrm{eq}}$ independently, which agrees well with simulation observations. These findings provide fundamental insight into surface structure design for controlling droplet hydrodynamics and the contact time during impact.

## 1. Introduction

The dynamic interactions between liquid droplets and solid surfaces are ubiquitously encountered in many important applications such as, inkjet printing, spray cooling, anti-icing, painting, and coating[1,2]. Droplet impact on solid surfaces has been an active area of research because understanding the physics underpinning the droplet-solid interactions is essential for these applications. For instance, the droplet spreading and recoiling during impact determines the resolutions of inkjet printing[3,4]. The droplet hydrodynamics is also important for maximizing the covered area and maintaining a uniform film thickness in painting[5]. How long the droplet remains in contact with the hot solid surfaces influences the cooling efficiency of spray cooling[6,7]. Moreover, reducing the contact time is critical for inhibiting ice nucleation on anti-icing surfaces[8–10]. Despite extensive studies, the transport and interfacial phenomena associated with droplet impact are so rich and have yet been fully understood.

For the impact of micron-size droplets in which the gravitational force can be neglected, the dynamics is governed by the interplay of inertia, capillarity, and viscous dissipation. Two important dimensionless numbers determine the relative dominance between the effects. The Weber number ($We = 2\rho_l R_o U_o^{\,2} / \sigma$) determines the relative importance of fluid's inertia to its surface tension where $\rho_l$ represents the droplet density, $R_o$ is the initial radius of the droplet, $U_o$ is the impact velocity and $\sigma$ is the surface tension. The Ohnesorge number ($Oh = \eta_l / \sqrt{2\rho_l R_o \sigma}$) relates the viscous forces to the surface tension force, where $\eta_l$ represents the dynamic viscosity of the liquid droplet. These two non-dimensional numbers determine the impact Reynolds number i.e., $Re = \sqrt{We}/Oh$ which reflects the importance of inertia to viscous effects ($Re = 2\rho_l R_o U_o / \eta_l$). The majority of previous studies focused on impacts at moderate and high Weber numbers[1,2]. However, in the inkjet technology, which is increasingly important for additive manufacturing of electronics and bioassays, microscale molten metal droplets or droplets with surfactants and macromolecules typically impinge on substrates at low Weber numbers[11–13]. The low Weber number regime (i.e., $We \leq 15$) displays an initial spreading phase dominated by the inertial force and a recoiling phase governed by the viscous and surface tension forces[14–16]. In addition to the liquid properties, the topography and wettability of solid surfaces also play a key role in determining the characteristics of impact dynamics and final outcomes[10,17–21].

For droplet impact on solid surfaces, various outcome scenarios, which range from deposition, partial or total rebound, to splashing, are influenced by droplet properties, kinematic of impact, and surface characteristics. In recent years, reducing the contact time between bouncing droplet and complex surface structures has attracted significant attention in the design of super-repellent surfaces and materials. Bange and Bhardwaj[22] numerically investigated the impact of droplet on superhydrophobic micro-textured surfaces and showed that the droplet bounces off the surface if the total droplet energy at the instance of maximum recoiling exceeds the initial surface and gravitational energy, otherwise not. A regime map was proposed to predict bouncing and non-bouncing on the superhydrophobic surface with an equilibrium contact angle of 155°. Li and Zhang[23] studied the dynamic behavior and impact outcomes of droplet on superhydrophobic surfaces via both experiments and molecular dynamic simulations. Impact outcomes including regular rebound, jetting, and splashing have been reported. The energy loss during impact was characterized by the restitution coefficient. More recently, Wang *el al*.[24] investigated the bouncing droplets on different micro- and nano-textured surfaces. They demonstrated a decrease in contact time of bouncing droplets on high solid fraction surfaces through reducing the texture size to nanometer scale. Theoretically, the authors showed that the contact time reduction can be attributed to the dominance of three-phase contact line tension on compact nanoscale textures.

Droplet impact on solid flat surfaces has been extensively investigated through experiments, theoretical modeling, and computer simulations[1,2]. In contrast, past studies on the dynamic behavior of droplet impact on curved surfaces are limited. Important issues regarding the physical properties of liquid droplet, surface characteristics (geometry, wettability, and roughness), target-to-droplet size ratio, and impact velocity are not fully understood yet. These parameters can significantly affect the impact dynamics and outcomes. Moreover, a comprehensive and quantitative analysis of energy budget during impact is very limited or sometime missing in previous studies. Hardalupas *et al*.[25] experimentally investigated the impingement of monodisperse droplets of a water-ethanol-glycerol solution on a spherical surface with diameter of the range of 0.8-1.3 mm. The size and velocity ranges of droplets are 160-230 µm and 6-13 m/s, respectively. Their results revealed the influence of impact kinematics (i.e., droplet momentum), spherical surface roughness, and curvature on the crown formation and propagation. The study mainly focused on impact outcomes in which droplets either deposit or re-atomize. The boundary between droplets deposition and re-atomization was identified by defining a critical factor dependent on the Reynolds (800 < $Re$ < 1750) and Ohnesorge numbers (0.008 < $Oh$ < 0.02)[25]. Bakshi *et al*.[26] experimentally and theoretically studied the impact of droplet on a spherical target. The impact Reynolds and Weber numbers were in the ranges of 780-5525 and 35-287, respectively, and the target-to-droplet size ratio was in the range of 1.23-8.8. The study measured the spatial and temporal variations of film thickness on the surface of the spherical target. Three distinct temporal phases of the film dynamics were observed in their experiments, namely the initial deformation phase, inertia dominated phase, and viscosity dominated phase. The results uncovered the effect of the Reynolds number and target-to-droplet size ratio on the dynamics of the film flow on the target surface. Theoretically, a simplified quasi-one-dimensional approach was used to model the flow on the spherical target and an analytical expression for the time-dependent film profile was obtained for the inviscid, inertia dominated phase of spreading. The evolution equation for the film thickness near the north pole of the spherical surface in the viscosity dominated phase was also obtained and solved[26]. Throughout their impact study, the surface wettability was defined either as highly nonwetting (water on stainless steel) or highly wetting (isopropanol on stainless steel). Charalampous and Hardalupas[27] examined head-on collisions between droplets and spherical particles. The Weber numbers were in the range of 92 < $We$ < 1015 and the Ohnesorge numbers were in range of 0.0070 < $Oh$ < 0.0089. The target-to-droplet size ratio ( $R^* = R_s/R_o$ ) were in the range of $1.8 < R^* < 11.1$. In addition to the conventional deposition and splashing regimes, a regime is observed in the intermediate region, where the droplet forms a stable crown, was observed. In contrast to these studies focused on moderate and high Weber numbers with droplet-to-target ratios greater than one ( $R^* > 1$ ), Banitabaei and Amirfazli[28] experimentally investigated the impact of a droplet onto a spherical particle with a target-to-droplet ratio less than one ( $R^* < 1$ ). The effects of a significantly wider range of Weber number variation (0.1 < $We$ < 1146) with three different surface wettability ( $\theta^{eq} = 70^o$ , $90^o$ , and $118^o$ ) on the impact dynamics was studied. They mainly focused on the formation of a thin liquid film, i.e., lamella, during the impact onto a hydrophobic sphere at high velocities.

Computationally, Yan-Peng and Huan-Ran[29] developed a direct numerical simulation that couples the level-set method with the interfacial cell immersed boundary method to model droplets impinging on a spherical surface. They investigated the impact behavior of droplet on solid spheres of different diameters ( $2 < R^* < 10$ ) at Weber numbers 1 < $We$ < 150 and Ohnesorge number $Oh$

= 0.0831. The droplets were observed to deposit on different spherical surfaces through oscillations and the results suggest that the spherical target size has a significant effect on the impact dynamics. More importantly, the simulation reported a breakage phenomenon in the center of the droplet during its first receding stage when the target is small. All impact simulations in this study[29] was limited to the surface with 90° equilibrium contact angle, i.e., the neutral wetting case. Mitra *et al*.[30] reported the impact of a subcooled droplet on a spherical surface ( $R^* \approx 0.85$ ) with high thermal conductivity. The impacts of droplets of three different liquids, namely water, isopropyl alcohol, and acetone were explored at different Weber numbers ($1 < We < 47$). They investigated the droplet shape evolution and surface wetting upon droplet impact on cold and hot spherical surfaces at 20° and 250°C, respectively. In addition, the droplet spreading patterns in cold condition and the film boiling regime were simulated using the volume of fluid method. Zhu *et al*.[31] developed an immersed boundary method to numerically investigate the dynamics of droplet impact on hydrophilic spherical targets at $\sim 10^3$ Reynolds numbers and the Weber numbers in the range of 25-400 with sphere-to-drop size ratios in the range of 2-5. With the aid of numerical simulations, Zhu *et al*.[31] identified the key regimes in the spreading and retraction and quantitatively evaluated the effect of sphere-to-drop size ratio on the impact dynamics. Using the level-set method, Khojasteh *et al*.[32] and Bordbar *et al*.[33] computationally investigated the impact dynamics of water droplet at low Weber number ($5 < We < 30$) on hydrophobic (125°) and super-hydrophobic (163°) substrates with both flat and curved geometries. The effect of Weber number, equilibrium contact angle, and surface curvature on the dynamics of the impact was reported. While in Khojasteh *et al*.[32] the sphere-to-droplet ratio was in the range of $2 < R^* < 4$ , Bordbar *et al*.[33] investigated a wider range of sphere-to-droplet size ratio ($1 < R^* < 8$ ). A theoretical model which is able to predict the maximum spreading of droplets on spherical targets was provided by Bordbar *et al*.[33] to obtain more detailed information on the spreading behavior of droplet on spherical target, Milacic *et al*.[34] employed a coupled immersed boundary and volume of fluid method to investigate the effect of the curvature on the spreading behavior of a droplet at low impact velocities and a small Laplace number. Liu *et al.*[35] studied droplet impact on spherical surface using the Volume of Fluid method with dynamic contact angle. Droplet dynamic behavior was investigated with various Weber numbers on spherical targets having different radii and surface wettability. Furthermore, they established a theoretical model based on energy budgeting to predict the effect of target curvature on the maximum spreading of the droplet. They conclude that the surface curvature has a significant effect on the maximum spreading factor when the sphere-to-droplet size ratio $R^* < 10$.

In recent years, the lattice Boltzmann method (LBM) has become one of the most widely used computational technique for simulating multiphase flows[36,37].Unlike conventional CFD methods, the major advantage of the LBM originates from microscopic perspective that recovers the macroscopic continuity and Navier-Stokes (N-S) equations from a system of linear partial differential equations. A macroscopic fluid flow is the result of the collective behavior of microscopic particles, which is mathematically described by the evolution of particle density distribution functions (PDFs). The macroscopic variables that describe the fluid flow, such as density, velocity, and pressure, can be obtained by evaluating the moments of PDFs[37]. Therefore, the discretization and numerical solution of the continuity and N-S equations are not required. When compared with conventional CFD methods, the pressure field can be obtained directly in the LBM (i.e., no need to solve Poisson equation which can be computationally expensive), LBM can be easily parallelized and handle complex geometry well (i.e., the no-slip boundary condition can be handled by bounce-back scheme). For modeling droplet-surface interactions, the diffuse interface regularizes the stress and pressure singularity at the contact line in a more rational manner with no special treatment such as ad hoc slip conditions[38,39].

However, the same as other numerical methods, multiphase LBM simulations suffer from several notable drawbacks such as restriction on the kinematic viscosity, interface thickness, low density ratio between phases, accuracy, and computational cost. Notably, many of these limitations have been mitigated or at least played down in some new LBM formulations such as entropic LBM and simplified LBM[40–42]. Mazloomi *et al.*[42] reported a novel thermodynamically consistent entropic LBM for multiphase flow with high density ratio and high stability. The novel entropic LBM has been well validated against experiments for head-on droplet impact and a stable lamella formation was observed. Another innovative LBM was presented by Chen *et al.*[41,43] to achieve third order accuracy by introducing virtual streaming nodes and decoupling the streaming distance from the mesh spacing. The simplified high-accuracy LBM can remarkably save virtual memory and facilitate implementation of physical boundary conditions.

Despite the growing popularity of the LBM in fluid flow simulations, there are limited LBM simulations that reported droplet impact on curved surfaces. Shen *et al*.[44] employed a two-dimensional (2D) pseudo-potential LBM to study the effect of Weber number and equilibrium contact angle during the droplet impact on a solid sphere at a single droplet-to-target size ratio. The equilibrium contact angle was varied through controlling the magnitude of attraction potential between the droplet and the surface nodes. The 2D LBM model by Shen *et al*.[44] has no density and viscosity contrast between the two fluid phases. A 2D multi-relaxation time (MRT) LBM developed by Zhang *et al*.[45] was employed to investigate the dynamics of droplet impact on a curved target. The model was able to handle a high-density ratio of 580 and hydrophilic and hydrophobic surfaces. The temporal variations of film thickness on the north pole of the target object were investigated at various Reynolds, Weber, and Galilei numbers. The effect of various equilibrium contact angle and side impact was also investigated in Zhang *et al*.[45] model. Another study by Zhang *et al*.[46] employed a 3D LBM that can simulate the impingement of a liquid droplet on spherical surfaces at high density ratio. They investigated the temporal variation of the film thickness on the spherical target surface. The effect of the droplet Reynolds number and the target-to-drop size ratio on the film flow dynamics was investigated.

Despite these documented studies on droplet impact on spherical targets, the interplay between the Weber number, target-to-droplet ratio, and surface wettability have not be systematically investigated. More importantly, quantitative characterization of contact time and rebound conditions on spherical targets has not been available. In this paper, we aim to fill this knowledge gap by focusing on the droplet morphological evolution, spreading and receding dynamics, impact outcomes, contact time, and energy transfer during impact. A wide range of surface wettability has been used. Furthermore, the energy transfer during impact for rebounded and non-rebounded droplets have been elucidated and compared. We integrated the LBM simulation and theoretical analysis to discover how the contact time before rebound and the rebound threshold can be precisely controlled by surface geometry, wettability, and Weber number. Droplet impact on curved surface is very common in industrial applications such as humid air or refrigerant hitting the blades of fans, compressors, or turbines, electrical cables, and inkjet printing on complex geometries. Thus, this study will contribute a fundamental understanding of droplet impact dynamics to the related fields. In the following section we briefly describe the implementation of the axisymmetric LBM to the present problem. Section 3 discusses the simulation results and the rebound model of droplet impact on spherical targets. Conclusions are drawn in Sec. 4.

## 2. Numerical Method

The LBM employed in this study is based on the free-energy formulation, in which the total free energy of the system is described as[47]

$$\psi_{\text{total}} = \int_V \left( E_0(C) + \frac{\kappa}{2} |\nabla C|^2 \right) dV + \int_S \left( \phi_0 - \phi_1 C_s + \phi_2 C_s^2 - \phi_3 C_s^3 \right) dS \tag{1}$$

The first term $E_0(C) = \beta C^2 (C-1)^2$ represents the bulk free energy. $C$ is the fluid composition defined as $C = (\rho - \rho_g)/(\rho_l - \rho_g)$, i.e., $C = 0$ in gas phase and $C = 1$ in liquid phase. $\beta$ is a constant determines the degree of immiscibility between the phases. The gradient term in the volume integral represents the energy penalty associated with the spatial variations in the fluid composition. Because $C$ only changes in the interfacial region between the phases, this term represents the interfacial free energy density. $\kappa$ is a constant parameter that can be related to the interfacial tension. The chemical potential is given as $\mu = \partial E_0/\partial C - \kappa \nabla^2 C$. The composition profile across the liquid-gas interface is dictated by $C(z) = 0.5 + 0.5\tanh(2z/\xi)$, where $z$ is the distance normal to the interface and $\xi$ quantifies the interfacial thickness. Given $\xi$ and $\beta$, the gradient parameter and the interfacial tension can be calculated respectively as $\kappa = \beta\xi^2/8$ and $\sigma = \sqrt{2\kappa\beta}/6$[48].

The surface integral in Eq. (1) represents the free energy contribution of the interaction between the fluid and the solid surface, where $C_s$ is the composition at the solid surface $S$ and $\phi_i$ with $i$ = 1, 2, 3 … are constant coefficients. We adopted a cubic wetting boundary condition[48,49] with $\phi_0 = \phi_1 = 0$, $\phi_2 = 1/2\phi_c$, and $\phi_3 = 1/3\phi_c$, in which $\phi_c = \Omega_c\sqrt{2\kappa\beta}$. $\Omega_c$ is a non-dimensional wetting potential determined by the equilibrium contact angle ($\theta^{eq}$) through Young's equation, $\cos\theta^{eq} = (\sigma_{sg} - \sigma_{sl})/\sigma_{lg} = -\Omega_c$, where $\sigma_{sg}$, $\sigma_{sl}$ and $\sigma_{lg}$ are the interfacial tensions for the solid-gas, solid-liquid, and liquid-gas interfaces, respectively.

An axisymmetric multiphase flow can be described by the Cahn-Hilliard, continuity and momentum equations as[50]

$$\frac{\partial C}{\partial t} + \left( u_z \frac{\partial C}{\partial z} + u_r \frac{\partial C}{\partial r} \right) + C\left( \frac{\partial u_z}{\partial z} + \frac{\partial u_r}{\partial r} \right) = M\left( \frac{\partial^2 \mu}{\partial z^2} + \frac{\partial^2 \mu}{\partial r^2} \right) + \frac{M}{r}\frac{\partial \mu}{\partial r} - \frac{C u_r}{r} \tag{2}$$

$$\frac{\partial \rho}{\partial t} + \left( u_z \frac{\partial \rho}{\partial z} + u_r \frac{\partial \rho}{\partial r} \right) + \rho\left( \frac{\partial u_z}{\partial z} + \frac{\partial u_r}{\partial r} \right) = -\frac{\rho u_r}{r} \tag{3}$$

$$\rho\left[\frac{\partial u_z}{\partial t}+u_z\frac{\partial u_z}{\partial z}+u_r\frac{\partial u_z}{\partial r}\right]=-\frac{\partial P}{\partial z}+\eta\left(\frac{\partial^2 u_z}{\partial z^2}+\frac{\partial^2 u_r}{\partial r^2}\right)+\frac{\eta}{r}\frac{\partial u_z}{\partial r}-\frac{u_z u_r}{r} \tag{4}$$

$$\rho\left[\frac{\partial u_r}{\partial t}+u_z\frac{\partial u_r}{\partial z}+u_r\frac{\partial u_r}{\partial r}\right]=-\frac{\partial P}{\partial r}+\eta\left(\frac{\partial^2 u_z}{\partial z^2}+\frac{\partial^2 u_z}{\partial r^2}\right)+\frac{\eta}{r}\frac{\partial u_r}{\partial r}-\frac{\eta u_r}{r^2}-\frac{u_r u_r}{r} \tag{5}$$

$\rho$ is the local density of fluid, which varies across the interface. $u_r$ and $u_z$ are fluid velocities in the radial and axial directions, respectively. $P$ is the hydrodynamic pressure, $\eta$ is the dynamic viscosity, and $M$ is a mobility constant that takes positive values. The governing equations (2-5) were discretized and solved numerically according to the scheme introduced by Lee *et al*.[48,49] (i.e., isotropic discretization) with additional forcing terms introduced for the formulations in the cylindrical coordinates[51–53]. The D2Q9 lattice velocity scheme was employed in this study, which represents two dimensions along the radial ($x$) and axial ($y$) directions and 9 discrete velocity directions.

Compared with full three-dimensional models, the axisymmetric model allows us to simulate larger droplets and provides better resolutions for the interfacial dynamics. Figure 1 shows a schematic plot of the computational domain of size 420×840 in the lattice units (*lu*). The center of the spherical target of radius $R_s$ was placed at the origin of the computational domain. The liquid droplet had a radius $R_o = 100$ and the impact axis coincided with the left boundary of the domain. The large droplet size was used to minimize the effects of diffuse interface with a finite thickness of 4 *lu* on droplet deformation and impact dynamics. The bounce-back boundary conditions[48,49] were imposed at the surface of the target and the bottom substrate. The curved solid surface was approximated by a staircase[54] representation with the bounce-back condition applied on the surface nodes. Notably, the calculation of the directional derivatives on the surface nodes requires information from the nodes inside the solid domain. The stress and pressure singularity at the contact line is naturally regulated by the diffuse interface in LBM[38,39]. We assumed the reflection symmetry along a directional derivative, i.e. $\phi(\mathbf{x}_s+\mathbf{e}_i\delta t)=\phi(\mathbf{x}_s-\mathbf{e}_i\delta t)$, where $\phi$ represents any macroscopic variable, $\mathbf{x}_s$ represents the solid node position, and $\mathbf{e}_i$ represents the lattice velocity in the $i$-th direction that points toward the interior solid node. For the left boundary of the domain, we imposed the symmetric bounce-back boundary conditions[53,55]. No-flux boundary conditions were imposed on the top and the right boundaries of the domain. More details of the simulation method can be found in our previous work[56].

## 3. Results and Discussion

In each impact simulation, the droplet was equilibrated for 5,000 lattice time steps. It then moved toward the spherical target with a uniform impact velocity ($U_o$). The density of the liquid droplet to the surrounding gas was $\rho_l/\rho_g = 786$ and the dynamic viscosity ratio was $\eta_l/\eta_g = 131$ . Both density and viscosity ratios are consistent with physical values in experiments. The interfacial tension, thickness, and the mobility were fixed at $\sigma = 0.01$, $\xi = 4$, and $M = 0.6667$, respectively. The simulations were performed by an in-house, serial C code and a typical impact simulation took about 20 to 30 hours on a 3.3-GHz processor.

To isolate the effects of Weber number, target-to-droplet size ratio, and surface wettability, we independently vary *We*, $R^*$, and $\theta^{eq}$ while all other impact parameters were kept constant. The droplet impacts were carried out at four different Weber numbers i.e., $We = 1, 5, 8, \text{and } 15$ while keeping the Ohnesorge number constant at $Oh = 0.025$. This low Weber number regime ensures that the interaction occurred exclusively between the droplet and curved surface, and do not involve the bottom flat substrate. The corresponding Reynolds number ranges from 40 to 155. The full parameter space of this study is summarized in Table S1. To quantitatively describe the impact dynamics, we measured the spread factor $D^*$, the dynamic contact angle $\theta^{dyn}$, and the contact line velocity $V_{CL}$. The dynamic contact angle was measured based on the fluid composition gradient at the spherical surface. The contact line velocity was obtained by tracking the interfacial position on the spherical surface at which the composition $C = 0.5$.

### 3.1 Model Validation

We first validated our simulation model of droplet impact on spherical targets against previous studies[30,57]. The reference case considers impact on a neutral spherical surface with a relative size ratio $R^*$ = 3.226 at Weber number *We* = 8. The droplet morphologies are comparable between experiments, two previous simulations, and current simulation, shown in Figure S1. Figure 2 further quantifies the impact dynamics using spreading factor defined as $D^* = L_w/D_o$, where $L_w$ is the instantaneous wetted length on the spherical target and $D_o$ is the initial droplet diameter before impact. The present simulation shows a good agreement with the experiment in the early stage of impact (spreading phase) and demonstrates a better accuracy when compared with the simulations performed by Mitra *et al.*[30] The small discrepancy from experiments during the receding phase is likely attributed to the higher *Oh* of 0.025 necessary to maintain simulation stability than the one in the experiment (*Oh* = 0.0021). In addition, the accuracy of the wetting boundary conditions on the spherical surface was validated by measuring the apparent contact angle of a stationary liquid droplet on the target, as shown in Figure S2. The apparent contact angle is defined as the angle between the tangent to the spherical target and the tangent to the fluid interface at the point of three-phase contact line, i.e., at $r = R_s$ and $C = 0.5$ where *r* represents the radial coordinate with its origin at the center of the spherical target. The fluid interface was reconstructed based on the fluid composition gradient. The apparent contact angles measured in the simulations also agree well with $\theta^{eq}$ set in the free-energy boundary condition.

### 3.2 Impact on Spherical Surfaces with Various Radii

We first investigated the effect of $R^*$ on the impact dynamics at the reference *We* of 5. $R^*$ was varied from 1.25 to 3.75 with the wettability of the spherical target being neutral, i.e., $\theta^{eq} = 90^o$. The minimal $R^*$ was selected to guarantee the droplet always contacts the surface with a well-defined, uniform curvature even at the highest *We* explored in this study. Figure 3 shows the sequences of velocity vectors and pressure fields during droplet impact on the spherical targets. The initial droplet deformations immediately after contact are approximately the same for all targets. A high-pressure zone developed in the lower droplet near the spherical target in all three cases. However, the pressure builds up much faster and the high-pressure zones are notably larger for impacts on $R^*$ = 1.5 and 2.25, which is likely due to the influence of surface curvature. The formation of liquid lamella around the target was clearly observed. The droplets continued to spread out until the inertia of the liquid droplet completely diminished. At this instant, the maximal spreading was attained. By the end of the primary spreading phase, the directions of the velocity vectors reversed and all droplets started to recede. Despite similar spreading dynamics, the droplets exhibited distinct receding behavior with different $R^*$. For instant, the faster receding behavior of the droplet on small $R^*$ results in a partial rebound. The magnitude of dynamic receding contact angle and receding contact line velocity determine the margin between partial and total rebound. According to the reports of Rioboo *et al.*[15] and Antonini[58], partial rebounds correspond to small dynamic receding contact angles while total rebounds were observed at large receding contact angles. Both cases of rebounds require enough kinetic energy possessed by the receding droplet. In contrast, no rebounds were observed for the impacts at $R^* = 2.25$ and $R^* = 3.0$. Instead, the droplets ultimately deposited after a few cycles of oscillations. While complete deposition was observed for impacts with $R^* = 2.25$ and $R^* = 3.0$, the evolution in velocity vectors and pressure fields at the same instant of non-dimensional time ($T = U_o t / D_o$) is not the same. For droplet impact on flat surfaces, it has been demonstrated that the time scale of droplet oscillation is related to droplet size, liquid density, and surface tension through $t \approx \sqrt{\rho D_o / \sigma}$ [59–62]. Our results show that the oscillation frequency for non-rebounded droplets impacting a curved surface is inversely related with $R^*$, which show a dependency of the receding dynamics on $R^*$. However, because of a high *Oh* number simulated in this study, the numbers of oscillation cycles in the deposited cases are not sufficient for characterizing the oscillation quantitively. The detailed analysis of the curvature effect on oscillation frequency is beyond the scope of the work.

The droplet deformation shown in Fig. 3 can be quantified by the spread factor $D^*$, which is shown in Fig. 4 with respect to a non-dimensional time, $T$. Here, the time before contact was excluded so that $T = 0$ corresponds to the instant of initial contact. At the early stage of impact, all droplets spread quickly until the maximal spreading was reached. The collapsed curves of $D^*$ indicate that the first spreading phase was dominated by the droplet inertia and displayed weak dependency on $R^*$. In contrast, the droplet recoiling was strongly dependent on the target size. Different $R^*$ resulted in distinct receding behaviors of droplets and therefore different impact outcomes, which can be clearly shown in the evolution of $D^*$. For the impacts at $R^* = 1.5$ and $R^* = 1.75$, partial rebound of droplets were observed, while all droplets with $R^* \geq 2.0$ showed alternating periods of spreading and receding and eventually deposited onto the targets. Figure 4 clearly demonstrates that the characteristics of the oscillations of $D^*$ before the deposition was controlled by $R^*$. The larger $R^*$ led to oscillations with smaller magnitudes and higher frequencies.

The spreading and receding of impact droplets were further characterized by the contact line dynamics. Figures 5 and 6 show the instantaneous dynamic contact angle ($\theta^{\mathrm{dyn}}$) and the corresponding contact line velocity ($V_{\mathrm{CL}}$) during impact, respectively. A quick reduction in $\theta^{\mathrm{dyn}}$ from 180° to the equilibrium value $\theta^{\mathrm{eq}} = 90^\circ$ was observed as a result of droplet wetting and the inertia of impact. The large positive $V_{\mathrm{CL}}$ indicates the rapid spreading. The contact angle reduction overshot below 90° due to the remaining inertia until $V_{\mathrm{CL}} = 0$. Simultaneously, the spread factor reached the maximum. Afterwards, the dynamic contact angles started to increase, and the contact line velocities turned negative, which indicate the receding of contact line. The recoiling droplets exhibited different contact line velocities for different $R^*$. For the impacts on small targets, slightly larger receding velocities produced partial rebounds. The contact line velocity remains negative and its magnitude even increases again during the thinning of the liquid column (see Fig. 6). For the impacts on large targets, both $\theta^{\mathrm{dyn}}$ and $V_{\mathrm{CL}}$ underwent damped oscillations, which was attributed to the alternation between the spreading and receding phases. The contact line velocities fluctuated between negative and positive values and eventually became zero as the contact angles of the deposited droplets relaxed toward the equilibrium value.

These results suggest that the curvature of the spherical target has small effect on the maximal spread factor and on the early spreading dynamics. In particular, the maximum spread factor increases slightly for the impacts at smaller $R^*$. In contrast, the receding dynamics is significantly different for different values of $R^*$. More importantly, the amount of time that the droplet is in contact with the surface clearly depends on the $R^*$ value. Reducing the contact time will be analyzed in more detail in Sec. 3.4. It has been shown that the bouncing of the droplet is enhanced by the amount of the surface energy a droplet possesses during retraction[63]. In Sec. 3.5, a theoretical analysis of droplet rebound criterion will be presented.

### 3.3 Impact on Spherical Surfaces with Different Wettability

In this section, we discuss the simulation results of droplet impact on the spherical targets with a constant $We = 5$, $R^* = 2.5$ and varying $\theta^{\mathrm{eq}}$ in the range of 50° to 145°, which represents a wide range of surface chemistry from hydrophilic to hydrophobic. As shown in Fig. 7, the difference in the droplet velocity vectors and pressure fields can be clearly observed shortly after the contact ($T$ = 0.337). For impacting on the hydrophobic target, notable high-pressure and low-pressure zones were established in the lower and upper portions of the droplet, respectively. The resulting pressure gradient quickly diminishes the internal flow in the radial direction and restricted the droplet spreading. In contrast, the droplet impacting on the hydrophilic surface exhibited several localized pressurized zones but without global gradients to resist the radial flow. At $T = 0.78$, the droplet on the hydrophobic surface already reached its maximum spreading while the ones on the neutral and hydrophilic surfaces still slowly spreads (Fig. 8). The significant negative pressure at the center of the droplet and the positive pressure near the lamella edge promotes receding on the hydrophobic surfaces. After $T = 1.301$, all droplets were in the receding phase. No oscillations were observed for the impact on the hydrophilic surface. The droplet on the neutral surface exhibited oscillatory behavior and finally deposited, while the fast receding of the droplet on the hydrophobic surface resulted in a total rebound. The deviation in the impact dynamics behavior with surface wettability can be elucidated by the quantitative study of spread factor $D^*$, dynamic contact angle $\theta^{\text{dyn}}$, and the velocity of the contact line during the impact $V_{\text{CL}}$.

Figure 8 shows the spread factor of droplet impact on spherical targets with various wettability and constant $R^*$. The maximal spread factor is determined by the surface wettability. We observed larger maximal spread factors at lower equilibrium contact angles. This relation is attributed to higher attraction force between the interface and the solid surface. Furthermore, the droplets spread for longer time when impacting on surfaces with high wettability, which can be attributed to the lack of global pressure gradients that resist the radial flow as discussed above. Interestingly, different receding behaviors were observed afterward the maximal spreading. For the impacts on hydrophobic surfaces with $\theta^{\text{eq}} = 110^{\text{o}}$ and 130°, the rapid receding resulted in partial and total rebounds respectively, while complete depositions were observed on neutral and hydrophilic surfaces ($\theta^{\text{eq}} \leq 90^{\text{o}}$).

As shown in Fig. 9, the dynamic contact angle decreases monotonically toward the equilibrium value on the surface with. Oscillations were observed for the impacts on the surfaces with $\theta^{\text{eq}} = 70^{\text{o}}$ and 90°. For the impact on hydrophobic surfaces, the higher values of the dynamic receding contact angle led to droplet rebound. Total or partial rebound of a droplet was only observed when the receded droplets possess enough energy to overcome the attraction force between the droplet and the solid surface. This can be confirmed by investigating the contact line velocity for impacts on hydrophilic, neutral, and hydrophobic surfaces (see Fig. 10). During the main spreading phase, the contact line velocity rapidly reduces toward zero until the maximal spreading was reached. At this point, part of the kinetic energy was lost due to viscous dissipation and the rest was stored in the deformed droplet as the surface energy. Since most of viscous dissipation in the vicinity of the contact area, an increase in the surface hydrophobicity, which constrains the spreading of the droplet, will reduce the viscous dissipation[34,64]. For the impact on hydrophobic surface ( $\theta^{\mathrm{eq}} = 130^{\circ}$ ), less kinetic energy is dissipated by viscosity. Hence, the deformed droplet possesses more energy for the receding phase consequently rebound occur. To summarize, the smaller the energy dissipated by viscosity during impacts on hydrophobic surfaces and the higher the dynamic receding angle leads to rebound[64,65].

### 3.4 Impact Outcomes and Contact Time

We summarize the effect of surface curvature and wettability on the contact time and the impact outcomes in Fig. 11. We observed five outcomes: deposition, partial rebound, partial rebound with satellite droplet(s), total rebound with satellite droplet(s), and total rebound. The surface in Fig. 11 represents the magnitude of contact time between the droplet and the spherical target before rebound. While the reduction in the contact time with surface hydrophobicity is well known from the literature[1,15,58], the effect of surface curvature has yet been thoroughly studied. For the impact on cylindrical surfaces, Andrew *et al.*[66] and Liu *et al.*[67] claimed that the enhancement of droplet rebound is associated with momentum imbalance between the axial and azimuthal directions during impact. Furthermore, Andrew *et al.*[66] found that the contact time is minimized when the target-to-droplet ratio is 1 and increases when the target-to-droplet ratio is greater or less than 1. Our result is consistent with these studies. The gradient of the contact time surface reveals the relative dominance of these two parameters. In particular, the effect of equilibrium contact angle is stronger. We believe that the reduction in contact time with smaller $R^*$ caused the faster receding of the droplet. Because the maximum spreading factor is almost independent of the $R^*$, this means higher curvature associated with smaller $R^*$ which produces higher surface energy available during the receding of the droplet, as a result the droplet rebound eventually.

We implemented nonlinear regression to predict the contact time in terms of target-to-droplet size ratio and surface wettability. The resulting model is

$$T_{\mathrm{CT}} = 4.244 - 1.08\,\theta^{\mathrm{eq}} + 0.545\,R^* - 0.2618\,\theta^{\mathrm{eq}}\,R^* \tag{6}$$

Even though the impact dynamics is clearly different, previous studies[68,69] indicate that the drop contact time $T_{CT}$ is independent of the Weber number ($We$ ~ 1-18), instead it scales with the inertial-capillary timescale ($T_{CT} \sim \sqrt{\rho D_o^{\ 3}/\sigma}$). For impact on spherical surface, Khojasteh *et al.*[32] indicated an identical contact time for impacts on hydrophobic surfaces (125°) with two different Weber numbers ($We$ = 5 and 15). Consequently, Eq. (6) provides a good model for predicting the contact time of rebounded droplets after impacting spherical surfaces for the current simulations range.

### 3.5 Theoretical Model of Droplet Rebound Criterion

To predict droplet rebound on spherical surface, we conducted a theoretical analysis based on total energy conservation during impact. At the maximum spreading, the droplet is assumed to be a thin liquid film with uniform height $h$ and wetted arc length $L_{w,\max} = 2R_s\varphi_{\max}$ as shown in Fig. 12, where $\varphi_{\max}$ (in rad) represents half of the central angle between the north pole of the spherical target and the three-phase contact line of the liquid film. A droplet is expected to rebound if the excess of surface energy at the maximum spreading state at least overcomes viscous dissipation and enables the droplet to recover its initial sphericity[70]. The normalized excess energy available for droplet rebound should be greater than zero, which can be written as

$$E^* = \frac{E_1 - SE_0^{\text{lg}} - SE_0^{sg} - VE_r}{SE_0^{\text{lg}}} \geq 0 \tag{7}$$

Here, $E_1 = \left(A_{\text{top}} + A_{\text{side}}\right)\sigma_{\text{lg}} + A_{\text{bottom}}\sigma_{\text{sl}} - A_{\text{bottom}}\sigma_{\text{sg}}$ represents the surface energy of the droplet at the maximum spreading, where, $A_{\text{top}} = 2\pi\left(R_s + h\right)^2\left(1-\cos\varphi_{\max}\right)$, $A_{\text{side}} = \pi\sin\varphi_{\max}\left[\left(R_s + h\right)^2 - R_s^2\right]$, and $A_{\text{bottom}} = 2\pi R_s^2\left(1-\cos\varphi_{\max}\right)$ are the top, side, and bottom areas of the droplet at the maximum spreading. $SE_0^{\text{lg}} = \pi D_0^2\sigma_{\text{lg}}$ is the surface energy of the rebounded droplet that recovers the spherical shape. $SE_0^{\text{sg}} = A_{\text{bottom}}\sigma_{\text{sg}}$ represents the surface energy recovered by the re-establishment of the solid-gas interface during receding and rebound. $VE_r$ is the amount of energy dissipated due to viscous friction during droplet receding. We approximate the receding viscous dissipation by the spreading viscous dissipation, which can be obtained as $\frac{\pi}{12}\eta U_0 L_{w,\max}^2\sqrt{Re}$ [70]. The normalized excess energy $E^*$ can be then written as a function of *We*, *Re*, maximum spread factor $D_{\max}^*$, target-to-droplet size ratio $R^*$, and surface wettability

$$E^* = S_1 + S_2 + S_3 - \frac{\left(D_{\max}^*\right)^2}{12}\frac{We}{\sqrt{Re}} - 1 \tag{8}$$

where $S_1$, $S_2$, $S_3$ are functions of $D_{\max}^*$, $R^*$, and $\theta^{\mathrm{eq}}$. The rebound requires $E^*$ to be greater than or equal to 0. A detailed derivation of $VE_r$ and $E^*$ is presented in the appendix. Figure 13 plots the excess energy available for rebound calculated using Eq. 8 with impact variables *We*, *Re*, $R^*$, $\theta^{\mathrm{eq}}$, and $D_{\max}^*$ obtained from simulations with different surface wettability and target-to-droplet ratios.

The impact outcomes observed in the simulations and predicted by the excess energy are compared in Fig. 13. Figure 13a shows the normalized excess energy versus surface wettability with $R^*$ = 2.25 for four different Weber numbers i.e., *We* = 1, 5, 8, and 15 while Fig. 13b shows excess energy versus $R^*$ at the same four Weber numbers for a neutral surface with $\theta^{\mathrm{eq}} = 90^{\circ}$. The effect of surface wettability on the excess energy is more obvious when increasing the Weber number. At the lower Weber number i.e., *We* = 1 the excess energy is negative even for hydrophobic surfaces. The reduction in $R^*$ clearly increases the excess energy available for rebound. However, the smallest Weber number of 1 results in no rebound in the entire range of $R^*$ considered in this work. The direct comparison shows excellent agreement between the LBM simulations and theoretical predictions without any fitting parameters. Tables S2-4 summarize the results from simulations and theoretical analysis based on excess energy for different combinations of Weber number, $R^*$, and surface wettability.

To provide additional insight into the rebound criteria determined by the energy analysis, Figure 14 plots the evolution of energy components during droplet impact on a hydrophobic surface 130° (rebounded droplet) and on a hydrophilic surface 50° (non-rebounded droplet) with a fixed Weber number *We* = 8 and $R^*$ = 2.25. For the rebounded droplet, the surface energy increases drastically during the spreading phase due to the large deformation of the droplet and establishment of the energetically unfavorable liquid-solid interface. The kinetic energy is minimized and the surface energy is maximized at the maximum spreading. After that when the droplet starts to recede on the hydrophobic surface, a rapid and notable gain in the kinetic energy from the conversion of the surface energy can be observed. This results in the rebound of the droplet. On the other hand, the non-rebounded droplet was not able to recover its kinetic energy from the surface energy due to significant viscous dissipation and the droplet slowly reached the equilibrium state. The total energies of both the rebounded and non-rebounded droplets are approximately conserved. The small discrepancy can be attributed to the deviations in the measured interfacial areas and other viscous dissipation mechanisms. Viscous dissipation in droplet impact is in general the sum of three contributions occurring in a precursor film, at the contact line, and in the core of the droplet[71]. Only the last one is considered in this analysis. We observe that the deviation is most pronounced when the contact line is rapidly advancing and receding. In addition, larger discrepancy occurs for the rebounded droplet, which is known to have much higher contact line velocity as shown in Fig. 10. This correlation indicates the contact-line dissipation has appreciable contribution in the energy balance.

## 4. Conclusions

An axisymmetric LBM was developed for simulating droplet impact on spherical surfaces with various target-to-droplet size ratios ( $R^*$ ), surfaces wettability ( $\theta^{eq}$ ), and Weber number. Different spreading dynamics and impact outcomes for droplet impacts on targets have been observed. The maximum spread factor increased with the reduction in $\theta^{eq}$ . On the other hand, the maximum spread factor was almost unaffected by the variations in $R^*$ . Remarkably, the receding phase displayed distinct behavior for impact with various $\theta^{eq}$ and $R^*$ . The fast receding of the droplet for impacts with small $R^*$ produced a partial or total rebound of the droplet. The fast receding of the droplet was also observed for impacts on hydrophobic surfaces which a total rebound was observed. An outcome phase map was created to provide a comprehensive view of the effect of $R^*$ and $\theta^{eq}$ at a specific Weber number. Five different outcomes were defined: deposition, partial rebound, partial rebound with satellite droplet(s), total rebound, and total rebound with satellite droplet(s). Interestingly, the phase diagram shows rebound can be observed for impacts on hydrophilic targets ( $\theta^{eq} = 85^\circ$ ) when $R^* \leq 1.25$ if the Weber number is large enough. On the other hand, for Weber number of 1 no rebound was observed on hydrophobic surfaces. Contact time was shown to be inversely related with $R^*$ through a statistical analysis. We further performed an energy analysis of impact on spherical targets to systematically elucidate the effects of $R^*$, surface wettability, and Weber numbers on rebound criterion. We have demonstrated that the rebound criterion can be well predicted by the theoretical model based on energy conservation considering the surface curvature and wettability. Our findings would provide a deeper understanding of the droplet impact on spherical surfaces, which will add an important contribution toward controlling droplet impact dynamics for inkjet printing, cooling, coating, anti-icing, and many other applications.

**Supplementary Material:** Supplementary figures for numerical model validation and theoretical analysis and supplementary tables summarizing the comparison between simulation observation and theoretical predictions of impact outcomes.

**Acknowledgments:** X.Y. acknowledges partial support from the National Science Foundation under Grant No. CMMI-1538090. Computational resources were provided by the Watson Data Center at Binghamton University.

**Data Availability Statements:** The data that support the findings of this study are available from the corresponding author upon reasonable request.

**Appendix**

$VE_s \approx \Phi V t_s$ approximates the amount of energy dissipated due to viscous friction during droplet spreading[70]. Here, $\Phi \approx \eta\left(\partial_y u\right)^2 \approx \eta\left(U_0/L_c\right)^2$ represents the viscous dissipation power per unit volume. $V$ is the volume of viscous fluid (representing only a partial droplet) given by $V \approx \left(\frac{\pi}{4} L_{w,\max}^2 L_c\right)$, with $L_c$ being the characteristics length approximated as $L_c = 2D_0 / \sqrt{Re}$. $t_s$ is the time required for the droplet to reach the maximum spreading. Below, we estimate $t_s$ using volume and mass conservation of the droplet between the initial and maximum spreading states. Conservation of volume of the droplet between the initial state and maximum spreading gives

$$\frac{4}{3}\pi\left(\frac{D_0}{2}\right)^3 = \frac{2}{3}\pi\left(1-\cos\varphi_{\max}\right)\left[\left(R_s+h\right)^3 - R_s^3\right] \quad \text{(A.1)}$$

For the liquid film at the maximum spreading, Eq (A.1) can be approximated as

$$\frac{4}{3}\pi\left(R_0\right)^3 \approx 2\pi R_s^2 h\left(1-\cos\varphi_{\max}\right) \tag{A.2}$$

by assuming $h << R_s$. The liquid film thickness can then be written as

$$h \approx \frac{2}{3}\frac{R_0^3}{R_s^2\left(1-\cos\varphi_{\max}\right)} \tag{A.3}$$

We further consider that liquid flows from the droplet shaped like a truncated sphere into the film through an area of diameter *d* with velocity $U_0$ as shown in Fig. 14 gives

$$\pi\sin\varphi\left[\left(R_s+h\right)^2-R_s^2\right]U_R=\pi d^2 U_0 \tag{A.4}$$

Assuming an average value $d \sim D_0/2$ [70] and negligible $h^2$, the droplet mass conservation during spreading can be approximated as

$$2U_R R_s h\sin\varphi \approx \pi R_0^2 U_0 \tag{A.5}$$

Note that $U_R \approx R_s\left(d\varphi/dt\right)$, the substitution of Eq. (A.3) into Eq. (A.5) gives

$$\frac{d\varphi}{dt}=\frac{3}{4}\frac{\left(1-\cos\varphi_{\max}\right)}{R_s R^*\sin\varphi}U_0 \tag{A.6}$$

Rearranging and integrating both sides give

$$\int_0^{\varphi_{\max}}\sin\varphi d\varphi=\frac{3}{4}\frac{U_0}{R_0}\int_0^{t_s}dt \tag{A.7}$$

which can be written as

$$1-\cos\varphi_{\max}=\frac{3}{4}\frac{\left(1-\cos\varphi_{\max}\right)}{R_0}U_0 t_s \tag{A.8}$$

As a result, the spreading time can be approximated as

$$t_s=\frac{4}{3}\frac{R_0}{U_0} \tag{A.9}$$

and the viscous dissipation during droplet receding can be written as

$$VE_r \approx VE_s \approx \frac{\pi}{12}\eta U_0 L_{w,\max}^2\sqrt{Re} \tag{A.10}$$

Using Young's equation $\sigma_{\lg}\cos\theta^{\rm eq}=\left(\sigma_{\rm sg}-\sigma_{\rm sl}\right)$, the normalized excess energy can be expressed as

$$E^{*}=\frac{\left(A_{\rm top}+A_{\rm side}\right)\sigma_{\lg}-A_{\rm bottom}\cos\theta^{\rm eq}\sigma_{\lg}-\dfrac{\pi}{12}\eta U_0 L_{w,\max}^2\sqrt{Re}-\pi D_0^2\sigma_{\lg}}{\pi D_0^2\sigma_{\lg}} \tag{A.11}$$

Substituting the areas gives

$$E^{*}=\left\{\begin{matrix}\left(2\pi\left(R_s+h\right)^2\left(1-\cos\varphi_{\max}\right)+\pi\sin\varphi_{\max}\left[\left(R_s+h\right)^2-R_s^2\right]\right)\sigma_{\lg}\\ -2\pi R_s^2\left(1-\cos\varphi_{\max}\right)\cos\theta^{\rm eq}\sigma_{\lg}-\dfrac{\pi}{12}\eta U_0 L_{w,\max}^2\sqrt{Re}-\pi D_0^2\sigma_{\lg}\end{matrix}\right\}\Big/\pi D_0^2\sigma_{\lg} \tag{A.12}$$

$$\begin{aligned}E^{*}=&\,2\left(\frac{R_s}{2R_0}+\frac{h}{2R_0}\right)^2\left(1-\cos\varphi_{\max}\right)\\&+\frac{1}{2}\sin\varphi_{\max}\left[\left(\frac{R_s}{2R_0}+\frac{h}{2R_0}\right)^2-\left(\frac{R_s}{2R_0}\right)^2\right]\\&-2\left(\frac{R_s}{2R_0}\right)^2\left(1-\cos\varphi_{\max}\right)\cos\theta^{\rm eq}-\frac{1}{12}\frac{\eta U_0}{\sigma_{\lg}}\frac{L_{w,\max}^2}{D_0^2}\sqrt{Re}-1\end{aligned} \tag{A.13}$$

The substitution of $R^{*}=R_s/R_0$, $h^{*}=h/D_0$, $D_{\max}^{*}=L_{w,\max}/D_0$, $\varphi_{\max}=D_{\max}^{*}/R^{*}$, and capillary number $Ca=\eta U_0/\sigma_{\lg}$ into Eq. (A.13) gives

$$\begin{aligned}E^{*}=&\,2\left(0.5R^{*}+h^{*}\right)^2\left[1-\cos\left(D_{\max}^{*}/R^{*}\right)\right]\\&+\frac{1}{2}\sin\left(D_{\max}^{*}/R^{*}\right)\left[\left(0.5R^{*}+h^{*}\right)^2-\left(0.5R^{*}\right)^2\right]\\&-2\left(0.5R^{*}\right)^2\left[1-\cos\left(D_{\max}^{*}/R^{*}\right)\right]\cos\theta^{\rm eq}-\frac{1}{12}Ca\left(D_{\max}^{*}\right)^2\sqrt{Re}-1\end{aligned} \tag{A.14}$$

In terms of *We* number instead, Eq. (A.14) becomes

$$\begin{aligned}E^{*}=&\,2\left(0.5R^{*}+h^{*}\right)^2\left[1-\cos\left(D_{\max}^{*}/R^{*}\right)\right]\\&+\frac{1}{2}\sin\left(D_{\max}^{*}/R^{*}\right)\left[\left(0.5R^{*}+h^{*}\right)^2-\left(0.5R^{*}\right)^2\right]\\&-2\left(0.5R^{*}\right)^2\left[1-\cos\left(D_{\max}^{*}/R^{*}\right)\right]\cos\theta^{\rm eq}-\frac{\left(D_{\max}^{*}\right)^2}{12}\frac{We}{\sqrt{Re}}-1\end{aligned} \tag{A.15}$$

The strict conservation of volume between gives

$$\frac{4}{3}\pi R_0^3 = \frac{2}{3}\pi\left(1-\cos\varphi_{\max}\right)\left[\left(R_s+h\right)^3 - R_s^3\right] \tag{A.16}$$

which yields

$$h^* = \left[\frac{1}{4\left(1-\cos\varphi_{\max}\right)} + \left(\frac{R^*}{2}\right)^3\right]^{1/3} - \frac{R^*}{2} \tag{A.17}$$

Finally, the substitution of $h^*$ into Eq. (A.15) gives the normalized excess energy $E^*$ as a function of *We*, *Re*, maximum spread factor $D^*_{\max}$, target-to-droplet size ratio $R^*$, and surface wettability as

$$E^* = S_1 + S_2 + S_3 - \frac{\left(D^*_{\max}\right)^2}{12}\frac{We}{\sqrt{Re}} - 1 \tag{A.18}$$

where $S_1$, $S_2$, and $S_3$ can be written respectively as

$$S_1 = 2\left(\frac{1}{4\left[1-\cos\left(D^*_{\max}/R^*\right)\right]} + \left(\frac{R^*}{2}\right)^3\right)^{2/3}\left[1-\cos\left(D^*_{\max}/R^*\right)\right] \tag{A.19}$$

$$S_2 = \frac{1}{2}\sin\left(D^*_{\max}/R^*\right)\left[\left(\frac{1}{4\left[1-\cos\left(D^*_{\max}/R^*\right)\right]} + \left(\frac{R^*}{2}\right)^3\right)^{2/3} - \left(0.5R^*\right)^2\right] \tag{A.20}$$

$$S_3 = -2\left(0.5R^*\right)^2\left[1-\cos\left(D^*_{\max}/R^*\right)\right]\cos\theta^{\mathrm{eq}} \tag{A.21}$$

**Nomenclature**

| | |
|---|---|
| $A_{\mathrm{top}}$ | Top area of the droplet at the maximum spreading |
| $A_{\mathrm{side}}$ | Side area of the droplet at the maximum spreading |
| $A_{\mathrm{bottom}}$ | Bottom area of the droplet at the maximum spreading |
| $C$ | Composition |
| $Ca$ | Capillary number |
| $D^*$ | Spread factor of droplet |
| $E^*$ | Normalized excess energy |
| $E_0(\mathrm{C})$ | Bulk free energy |
| $\mathbf{e}_i$ | Lattice velocity in the *i*-th direction |
| $h$ | Liquid film thickness |
| $L_{\mathrm{C}}$ | Characteristics length |
| $L_w$ | Instantaneous wetted length |
| $M$ | Mobility |
| $Oh$ | Ohnesorge number |
| $P$ | Pressure |
| $R^*$ | Target-to-droplet size ratio |
| $R_o$ | Initial radius of the droplet |
| $R_s$ | Solid surface radius |
| $Re$ | Reynolds number |
| $SE_0^{lg}$ | Surface energy of the rebound droplet that recovers the spherical shape. |
| $SE_0^{sg}$ | Surface energy recovered by the reestablishment of the solid-gas interface during receding and rebound |
| $T$ | Non-dimensional time |
| $T_{CT}$ | Contact time |
| $U_o$ | Impact velocity |
| $V_{CL}$ | Contact line velocity |
| $VE_r$ | Amount of energy dissipated due to viscous friction. |
| $We$ | Weber number |
| $\mathbf{x}_s$ | Solid node position |
| $\beta$ | Immiscibility constant |
| $\eta_l$ | Droplet dynamic viscosity |
| $\theta^{eq}$ | Equilibrium contact angle |
| $\theta^{dyn}$ | Dynamic contact angle |
| $\kappa$ | Constant parameter related to interfacial tension |
| $\xi$ | Interfacial thickness |
| $\rho_l$ | Droplet density |
| $\sigma$ | Surface tension |
| $\Phi$ | Viscous dissipation power per unit volume |
| $\varphi_{max}$ | Half of the central angle between the north pole of the spherical target and the three-phase contact line of the liquid film. |
| $\phi_i$ | Constants in wetting free energy |
| $\phi_c$ | Constant related to wetting potential |
| $\phi$ | Any macroscopic property |
| $\Psi_{total}$ | Total free energy of a system |
| $\Omega_c$ | non-dimensional wetting potential |

**Figure Captions**

Figure (1): Schematic plot of the computational domain of size 420×840 for the axisymmetric impact of droplet on a spherical target.

Figure (2): Temporal evolution of spread factor ($D^*$) of droplet compared to experimental data and simulation results taken from Mitra *et al*.[30] for droplet impact at $We = 8$, $\theta^{eq} = 90^o$, and $R^* = 3.226$.

Figure (3): Velocity vectors and pressure fields during during impact on spherical solid surfaces with $We = 5$, $Oh = 0.025$, and equilibrium contact angle 90°. The ratios of the spherical surface radii to the droplet radius are $R^* = 1.5$ (left column), 2.25 (middle column), and 3.0 (right column).

Figure (4): Evolution of the spread factor after droplet impact on neutral spherical surfaces with various $R^*$. The asterisks mark the instant when the liquid bridge between the rebound droplet and the residual droplet ruptured.

Figure (5): Variations in the dynamic contact angles after droplet impact on neutral spherical surfaces with various $R^*$.

Figure (6): Instantaneous contact line velocities of droplets after impact on neutral spherical surfaces for three representative $R^*$. The dashed line marks the zero velocity.

Figure (7): Velocity vectors and pressure fields during the impacts on spherical solid surfaces with $R^* = 2.5$. The equilibrium contact angles are 50° (left column), 90° (middle column) and 130° (right column).

Figure (8): Evolution of the spread factor after droplet impact on spherical surfaces of $R^* = 2.5$ with different $\theta^{eq}$. The asterisks mark the instant when the rebounding droplet left the surface.

Figure (9): Variations in the dynamic contact angles after droplet impact on spherical surfaces of $R^*$ = 2.5 with different $\theta^{eq}$.

Figure (10): Instantaneous contact line velocities after droplet impact on spherical surfaces of $R^*$ = 2.5 with three representative $\theta^{eq}$. The dashed line marks the zero velocity.

Figure (11): Total contact time and impact outcomes: deposition (red star), partial rebound (black circle), partial rebound with satellite droplet (blue square), total rebound with satellite droplet (cyan diamond), and total rebound (magenta triangle). For deposition, the contact time is considered infinite.

Figure (12): Schematics of the initial and maximum spreading states during droplet impact on a spherical target.

Figure (13): (a) Normalized excess energy ($E^*$) versus surface wettability ($\theta^{eq}$) and (b) target-to-droplet ratio ($R^*$). The parameter regions of rebound and no-rebound are marked in light red and blue, respectively. The solid vertical line represents the rebound threshold defined by $E^* = 0$.

Figure (14): Time evolution of kinetic energy (circles), surface energy (squares), and accumulated viscous dissipation (upper triangles) normalized by the total initial energy for rebounded (black) and non-rebounded (blue) droplets at $We$ = 8, $R^*$ = 2.25 and $\theta^{eq}$ = 50° and 130° respectively. The total energies including these three components are plotted in green and magenta for rebounded and non-rebounded droplets, respectively. $T$ = 0 represents the instance when the droplet establishes contact with the target surface.

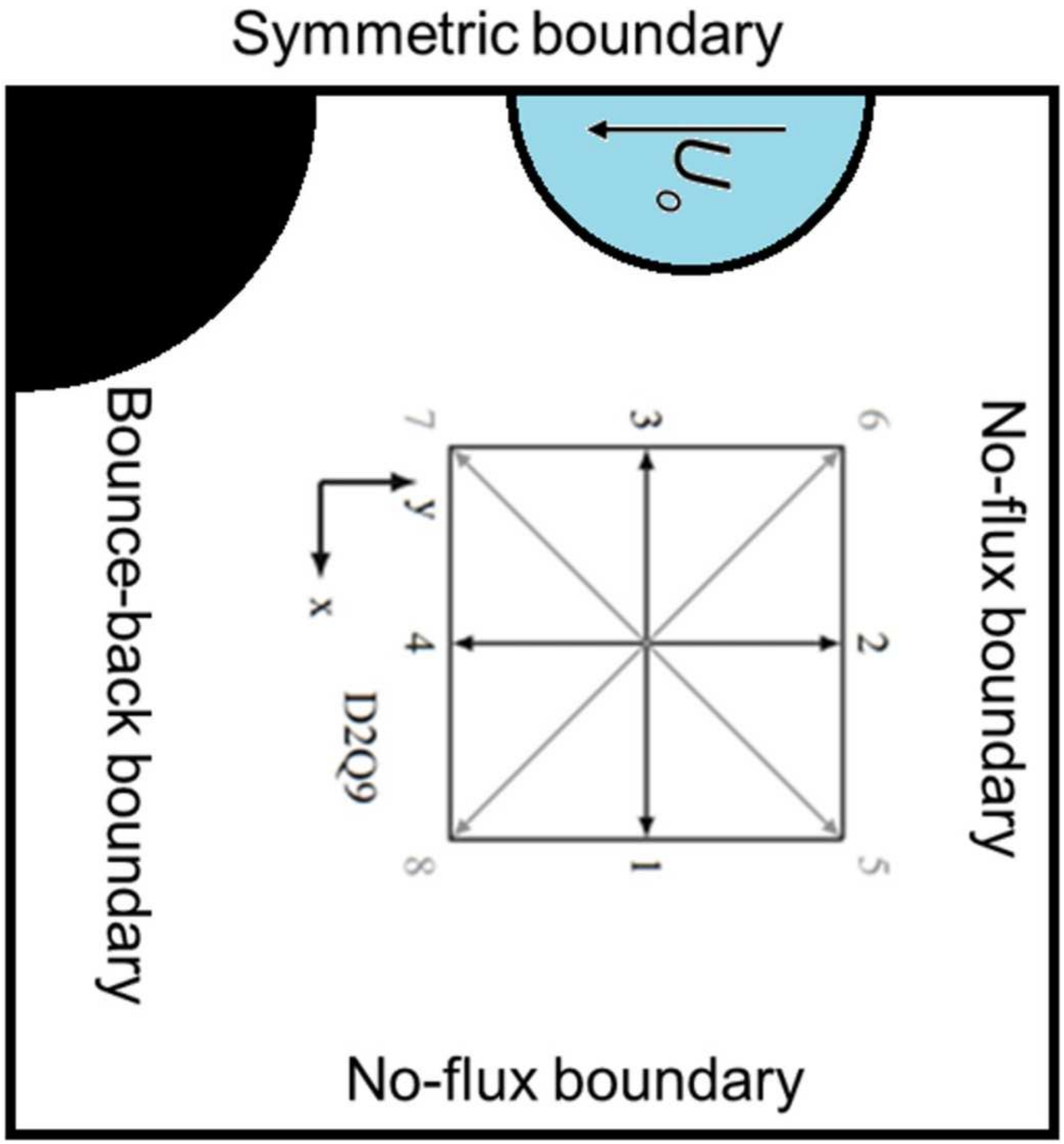
Symmetric boundary
$U_o$
No-flux boundary
No-flux boundary
Bounce-back boundary
1
2
3
4
5
6
7
8
y
x
D2Q9

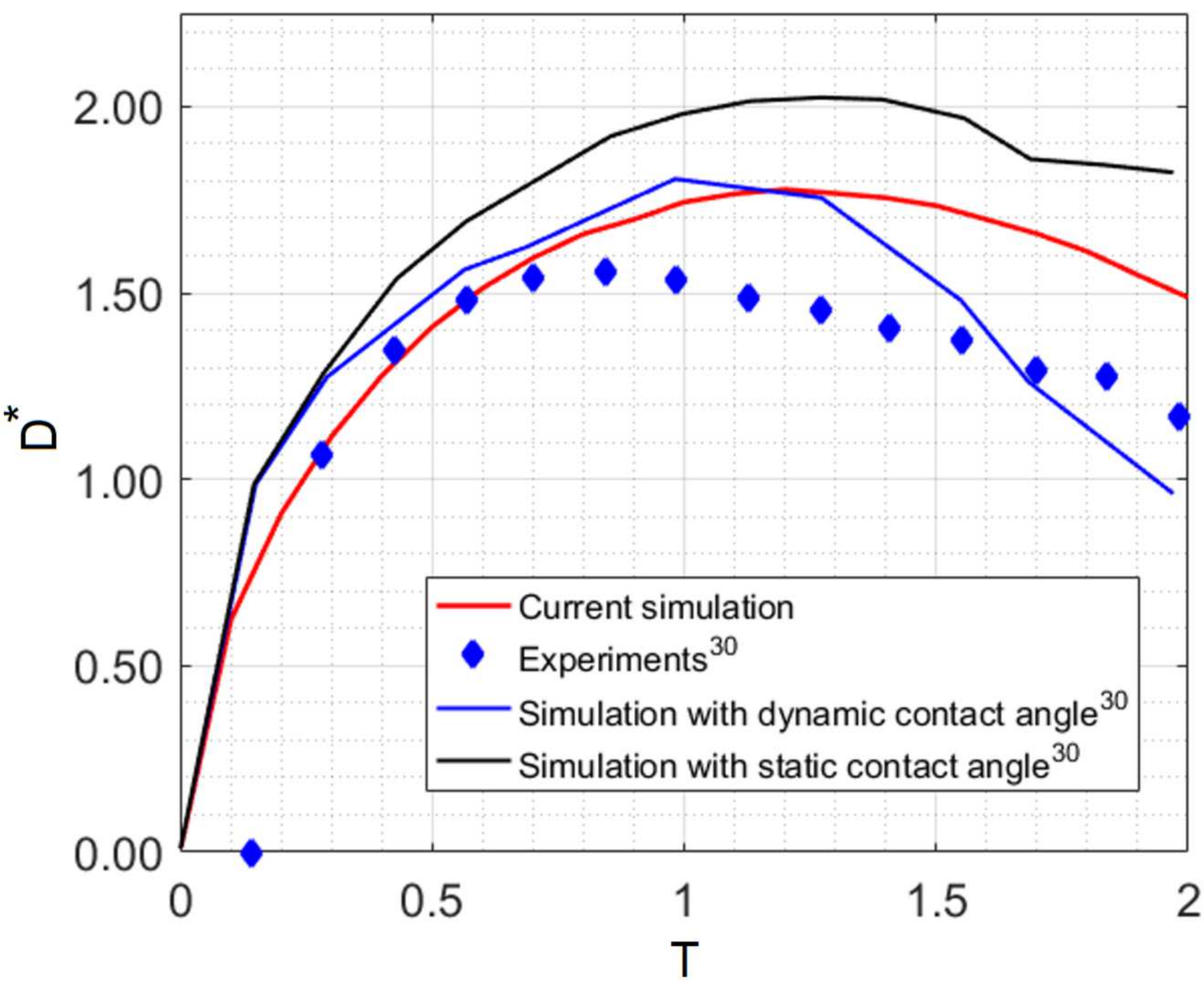
Current simulation
Experiments30
Simulation with dynamic contact angle30
Simulation with static contact angle30
D*
T
0.00
0.50
1.00
1.50
2.00
0
0.5
1
1.5
2

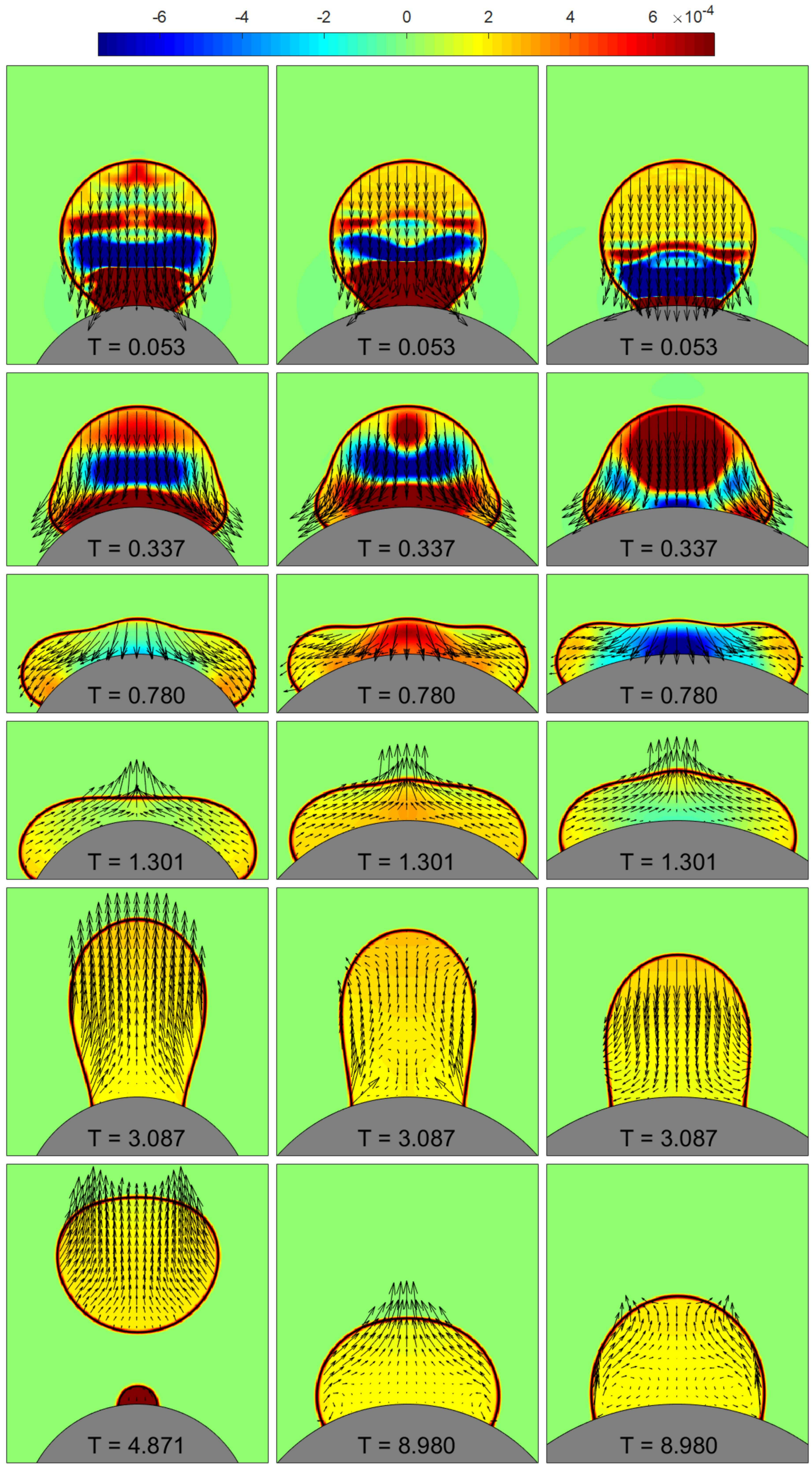
-6
-4
-2
0
2
4
6
$\times 10^{-4}$
T = 0.053
T = 0.053
T = 0.053
T = 0.337
T = 0.337
T = 0.337
T = 0.780
T = 0.780
T = 0.780
T = 1.301
T = 1.301
T = 1.301
T = 3.087
T = 3.087
T = 3.087
T = 4.871
T = 8.980
T = 8.980

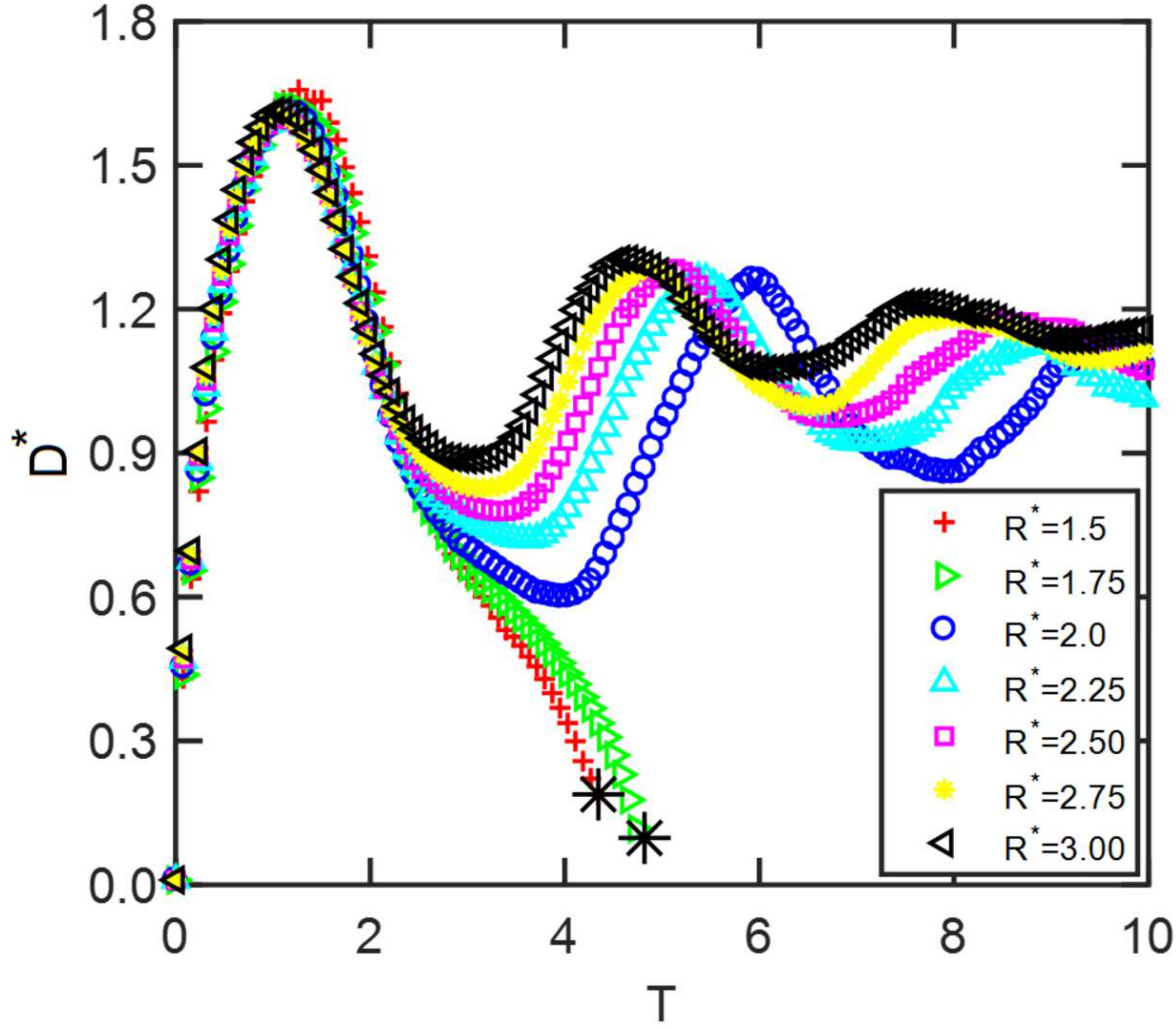
1.8
1.5
1.2
0.9
0.6
0.3
0.0
D*
0
2
4
6
8
10
T
R*=1.5
R*=1.75
R*=2.0
R*=2.25
R*=2.50
R*=2.75
R*=3.00

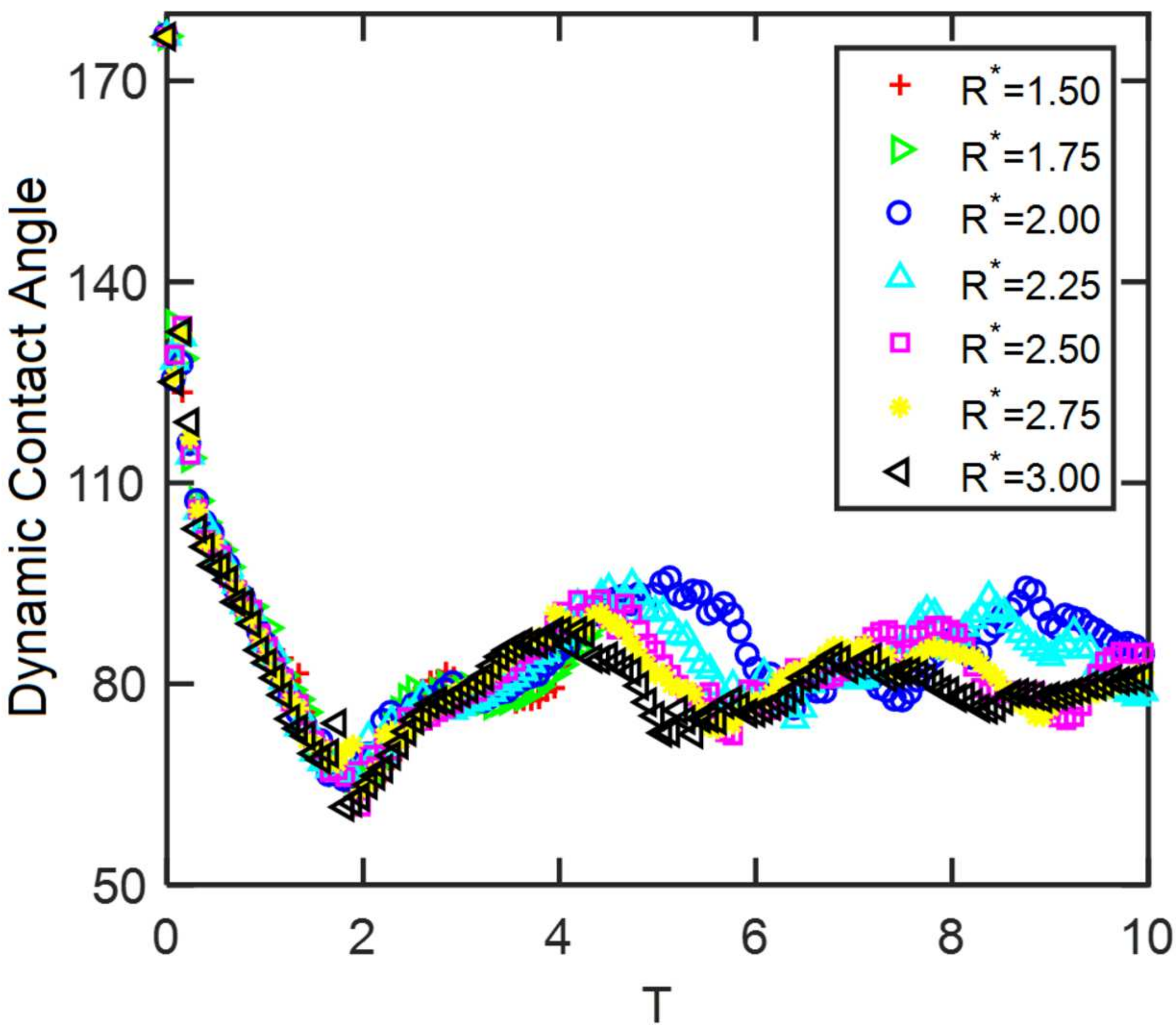
Dynamic Contact Angle
T
170
140
110
80
50
0
2
4
6
8
10
R*=1.50
R*=1.75
R*=2.00
R*=2.25
R*=2.50
R*=2.75
R*=3.00

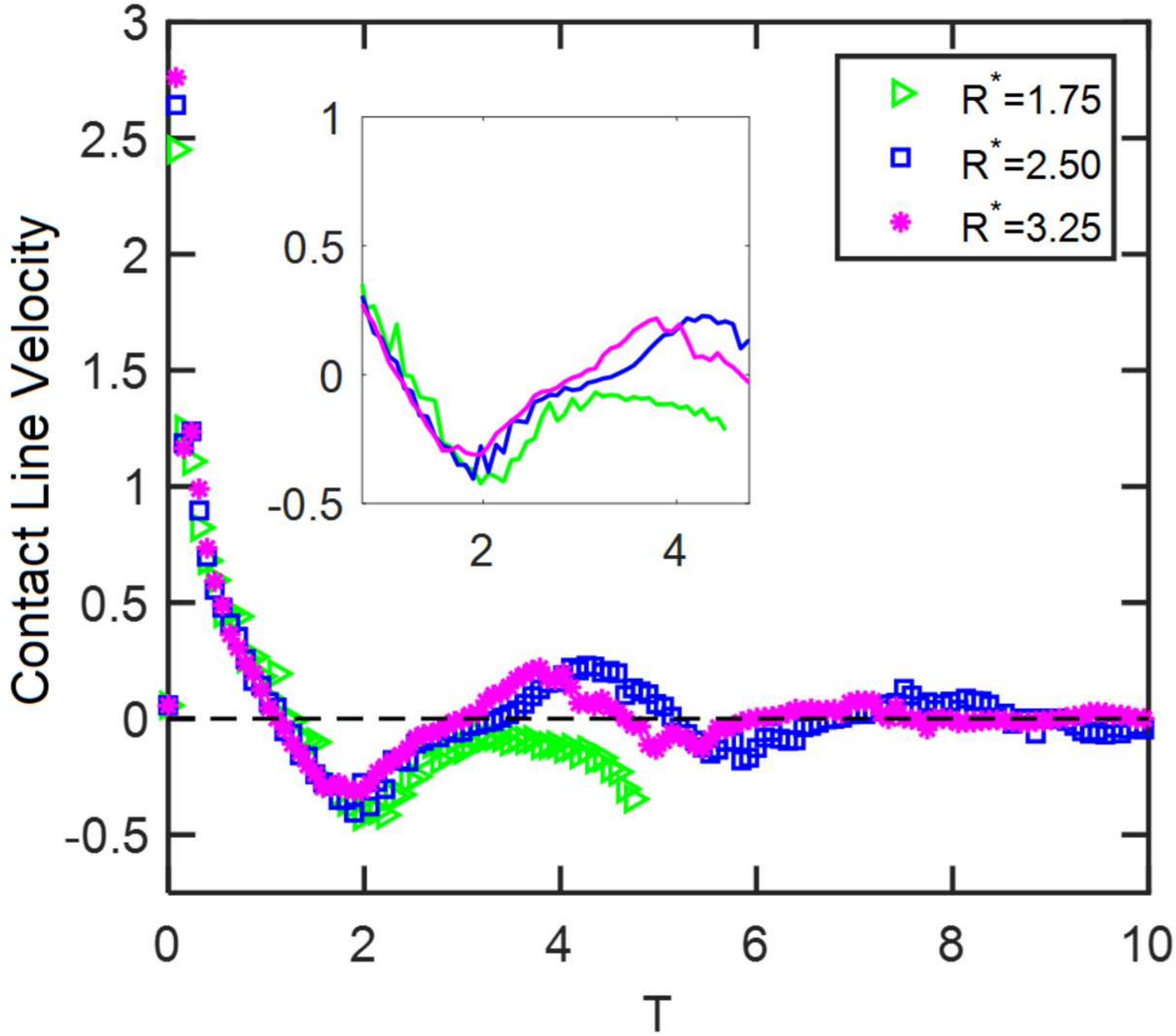
Contact Line Velocity
T
R*=1.75
R*=2.50
R*=3.25
0
2
4
6
8
10
-0.5
0
0.5
1
1.5
2
2.5
3

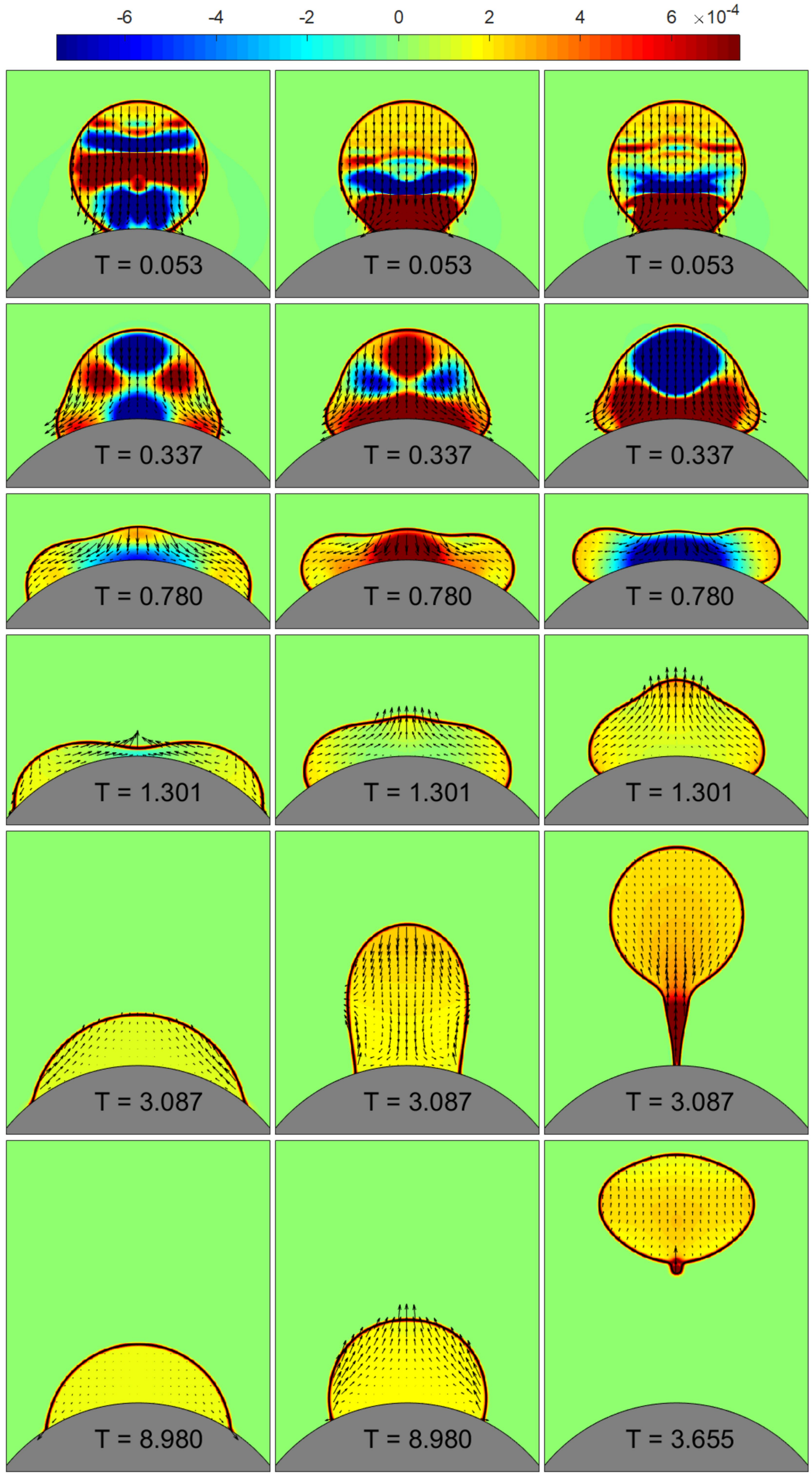
-6
-4
-2
0
2
4
6
×10^-4
T = 0.053
T = 0.053
T = 0.053
T = 0.337
T = 0.337
T = 0.337
T = 0.780
T = 0.780
T = 0.780
T = 1.301
T = 1.301
T = 1.301
T = 3.087
T = 3.087
T = 3.087
T = 8.980
T = 8.980
T = 3.655

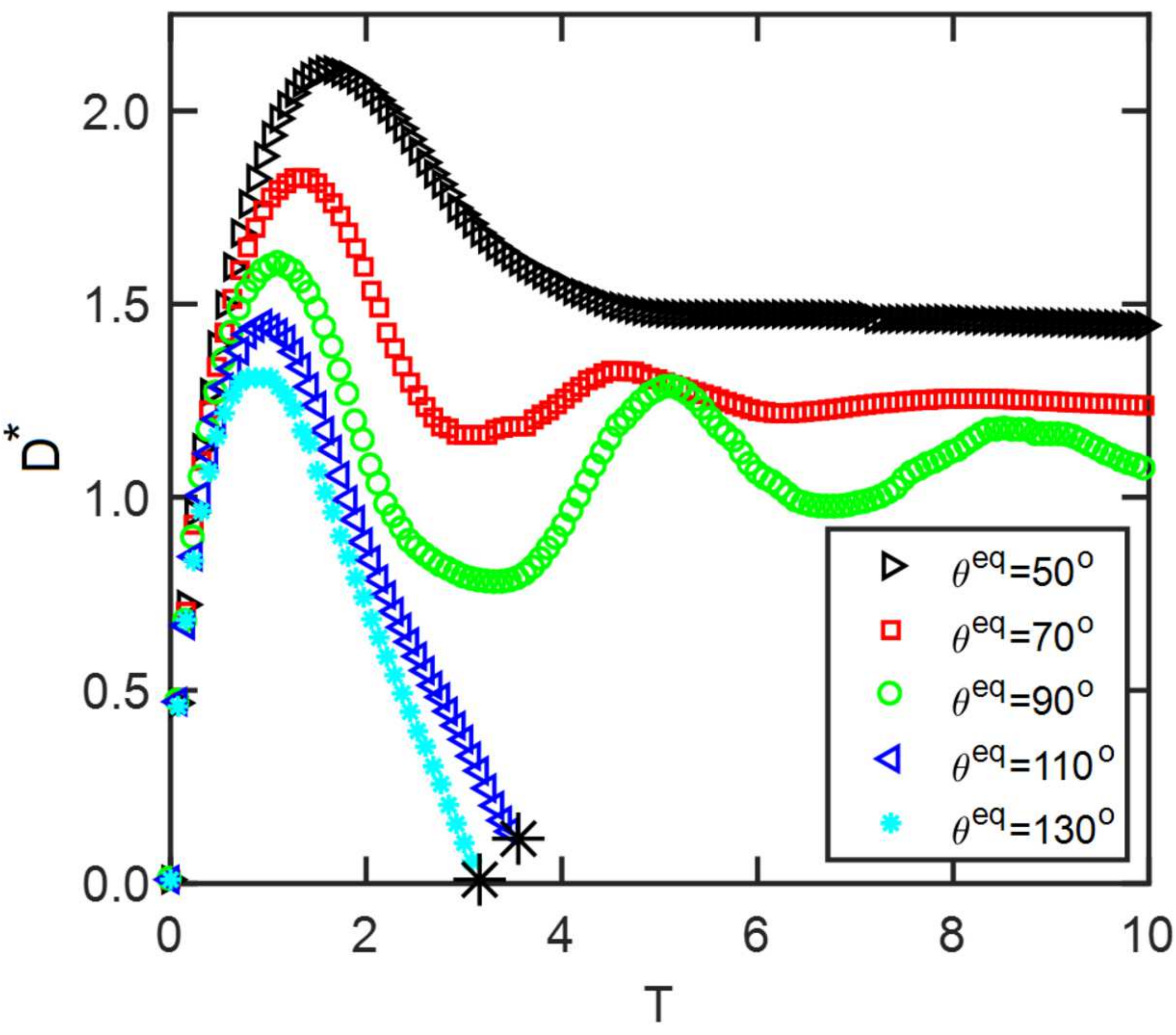
D*
T
0
2
4
6
8
10
0.0
0.5
1.0
1.5
2.0
θeq=50o
θeq=70o
θeq=90o
θeq=110o
θeq=130o

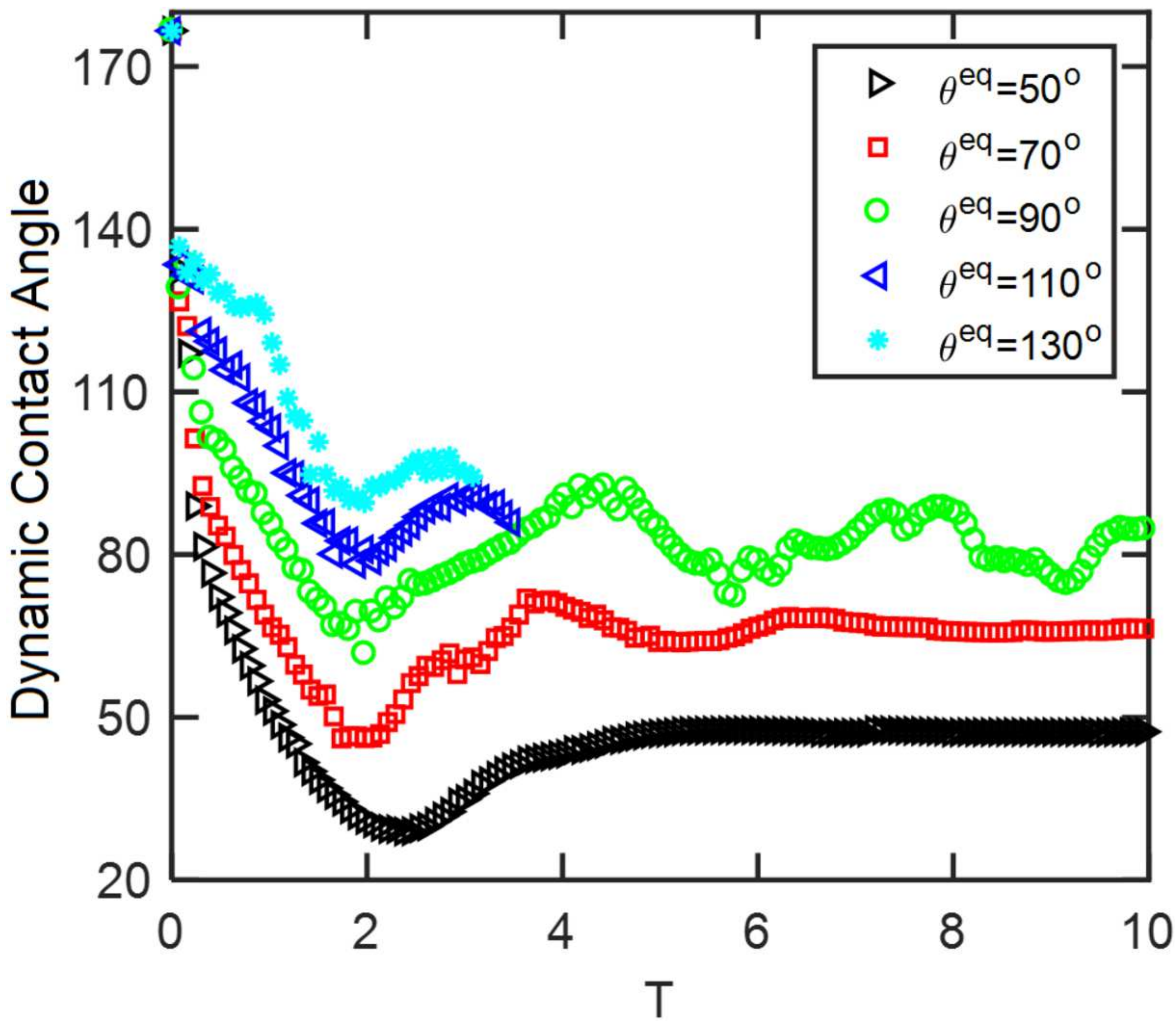
Dynamic Contact Angle
T
20
50
80
110
140
170
0
2
4
6
8
10
$\theta^{eq}$=50$^o$
$\theta^{eq}$=70$^o$
$\theta^{eq}$=90$^o$
$\theta^{eq}$=110$^o$
$\theta^{eq}$=130$^o$

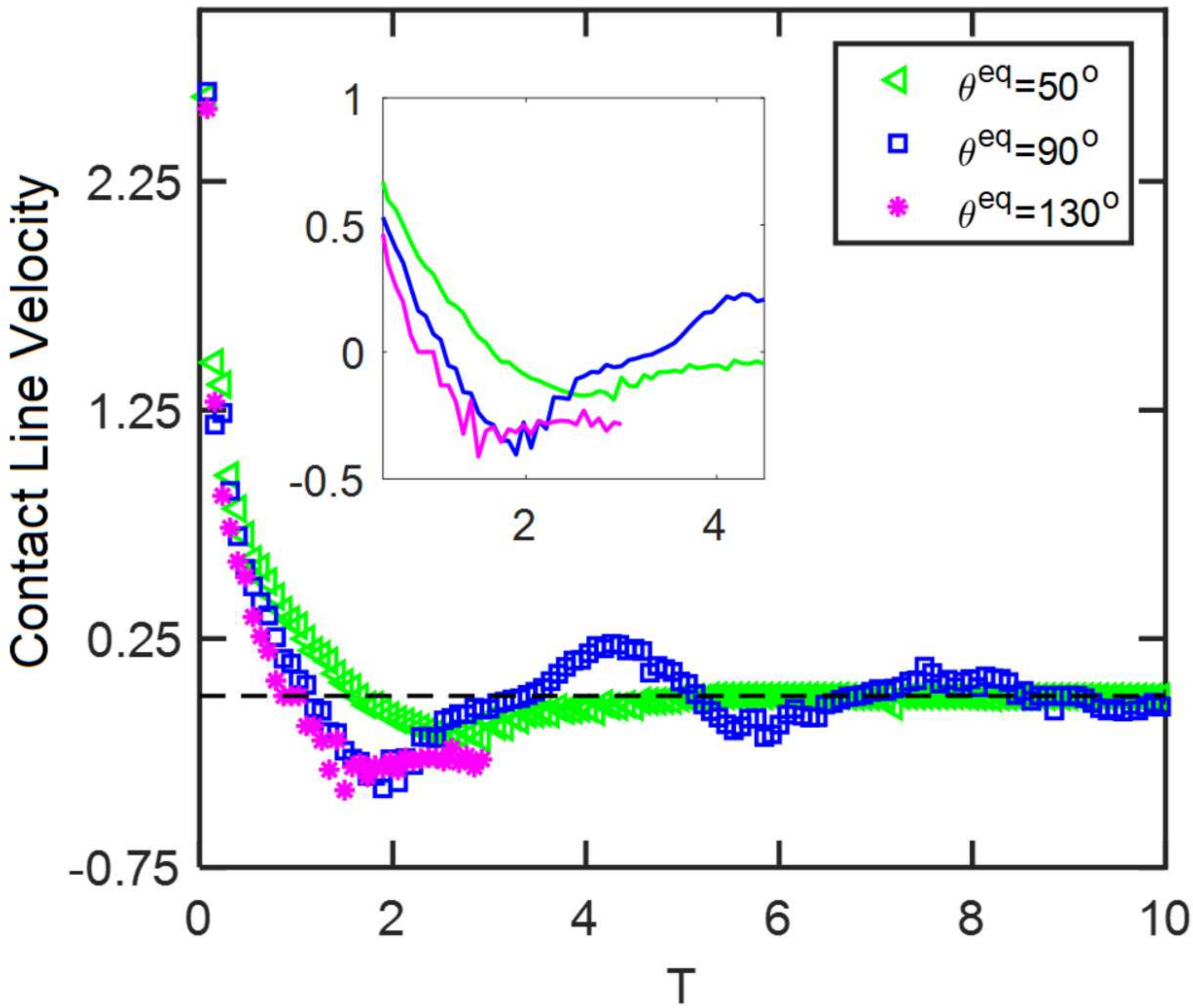
Contact Line Velocity
T
2.25
1.25
0.25
-0.75
0
2
4
6
8
10
$\theta^{eq}$=50°
$\theta^{eq}$=90°
$\theta^{eq}$=130°
1
0.5
0
-0.5
2
4

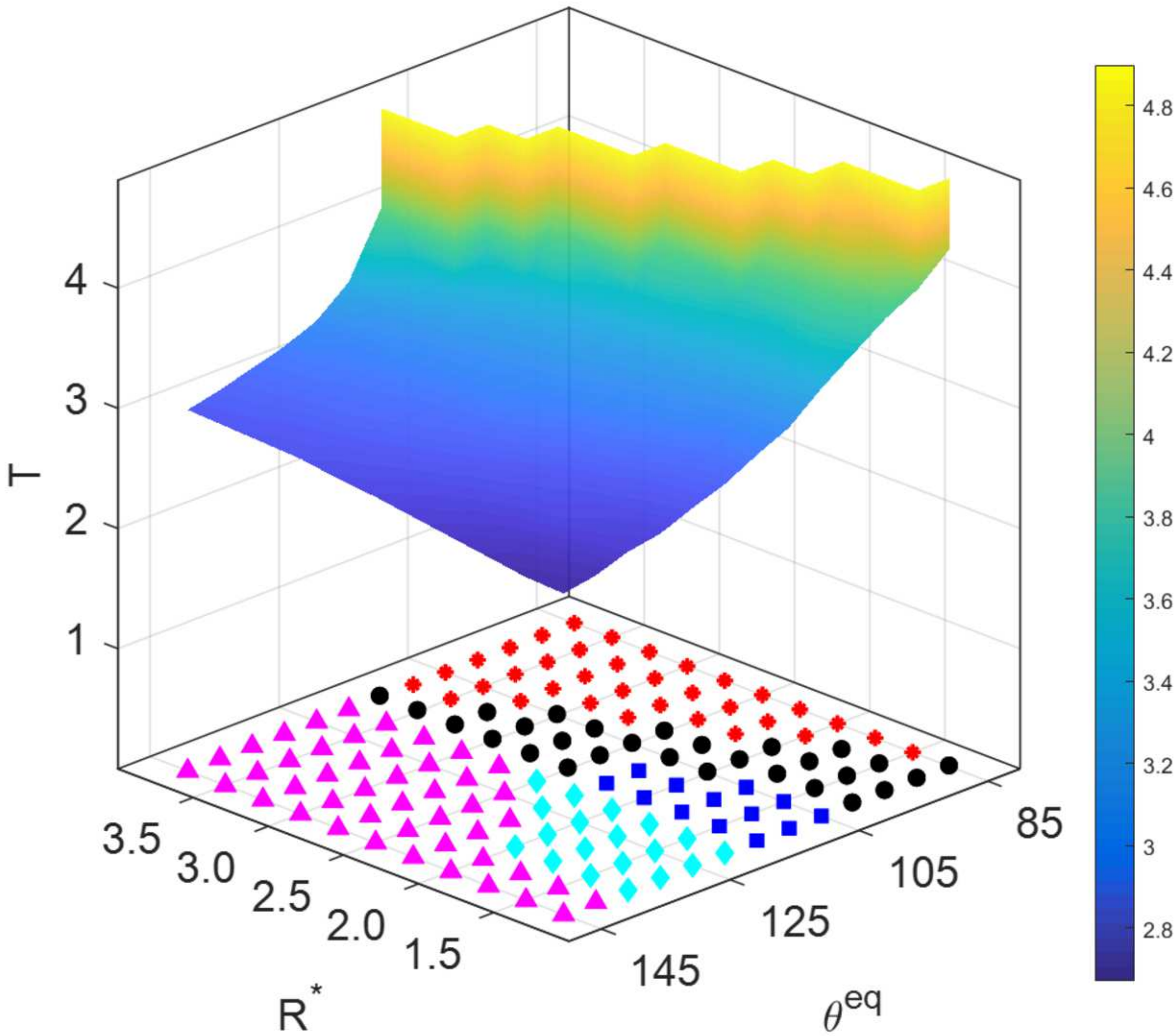
T
1
2
3
4
R*
3.5
3.0
2.5
2.0
1.5
$\theta^{eq}$
85
105
125
145
4.8
4.6
4.4
4.2
4
3.8
3.6
3.4
3.2
3
2.8

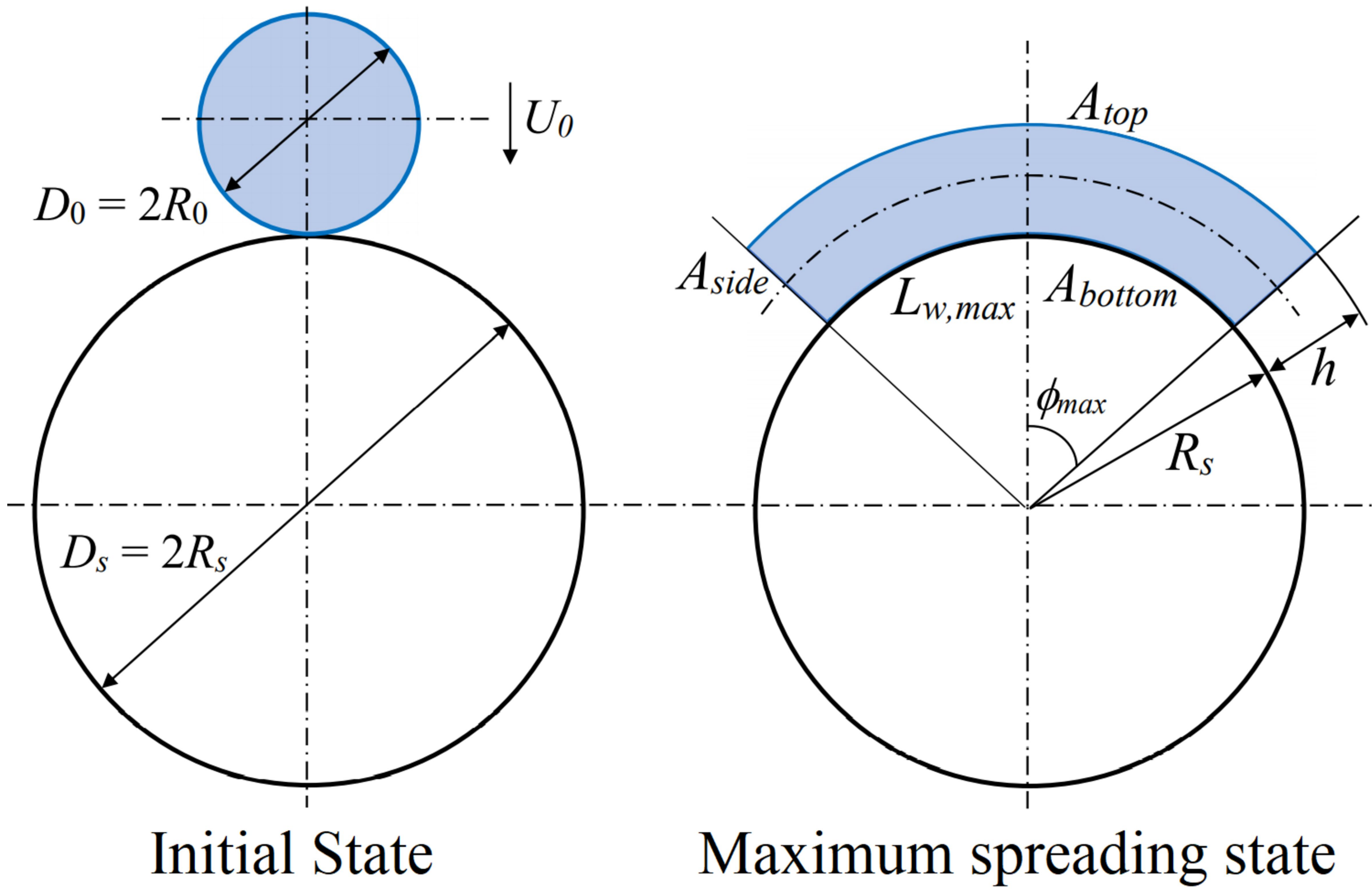
U0
D0 = 2R0
Ds = 2Rs
Atop
Aside
Lw,max
Abottom
h
ϕmax
Rs
Initial State
Maximum spreading state

Figure (13a).tif

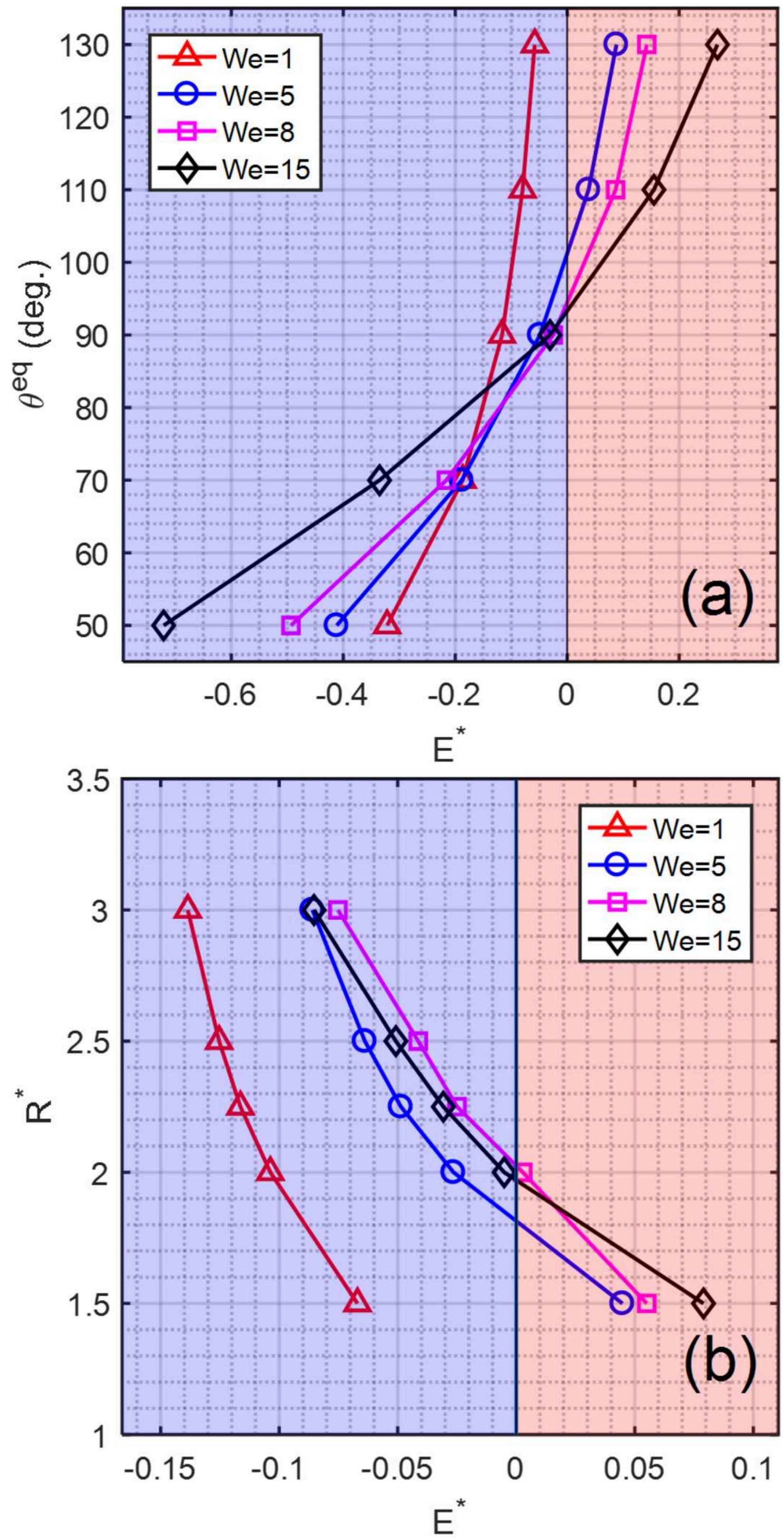

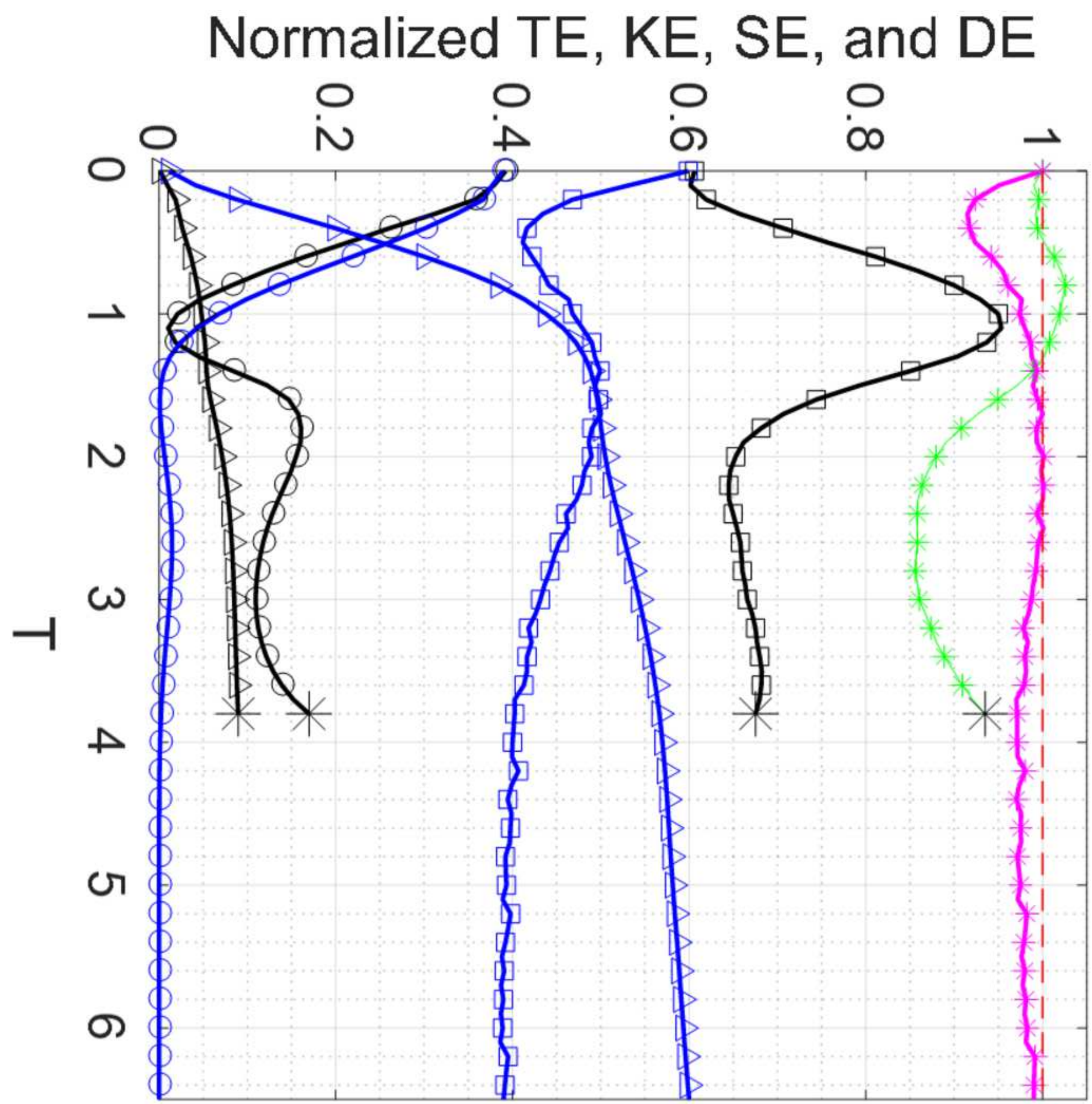
Normalized TE, KE, SE, and DE
0
0.2
0.4
0.6
0.8
1
0
1
2
3
4
5
6
T